\documentclass[10pt,twoside,twocolumn]{IEEEtran}

\usepackage{cite}
\usepackage[T1]{fontenc}
\usepackage{graphicx}
\usepackage{amssymb}
\usepackage{amsmath}
\usepackage{amsthm}
\usepackage{subfigure}
\usepackage{booktabs} 
\usepackage{multirow}
\usepackage{microtype}
\usepackage{balance}
\usepackage{xcolor}
\usepackage{xfrac}    
\usepackage{algorithm}
\usepackage{setspace}
\usepackage{mdframed}
\mdfdefinestyle{csstyle}{backgroundcolor=gray!10}
\usepackage{tikz}
\usepackage{circledsteps}
\usepackage{enumitem,kantlipsum}
\usepackage{dsfont}

\usepackage{algorithmicx}
\usepackage{algpseudocode}
\algrenewcommand\algorithmicindent{0.7em}%
\usepackage{xpatch}

\definecolor{myred}{HTML}{9E292B}
\definecolor{myblue}{HTML}{235787}
\definecolor{mygreen}{HTML}{5E6638}
\definecolor{mygray}{HTML}{444444}
\definecolor{myblack}{HTML}{000000}
\definecolor{mywhite}{HTML}{FFFFFF}

\definecolor{myaltred}{HTML}{D46A78}
\definecolor{myaltblue}{HTML}{6699C2}
\definecolor{myaltgreen}{HTML}{B0B58C}
\definecolor{myaltgray}{HTML}{AAAAAA}

\definecolor{mylightred1}{HTML}{B15455}
\definecolor{mylightred2}{HTML}{C57F80}
\definecolor{mylightred3}{HTML}{D8A9AA}
\definecolor{mylightred4}{HTML}{ECD4D5}
\definecolor{mylightblue1}{HTML}{5A7DA5}
\definecolor{mylightblue2}{HTML}{7D99BA}
\definecolor{mylightblue3}{HTML}{B3C3D7}
\definecolor{mylightblue4}{HTML}{D3DCE8}
\definecolor{mydarkgreen}{HTML}{3E4822}
\definecolor{mylightgreen1}{HTML}{828859}
\definecolor{mylightgreen2}{HTML}{9AA075}
\definecolor{mylightgreen3}{HTML}{B8BC96}
\definecolor{mylightgreen4}{HTML}{D4D4B8}
\definecolor{mylightgray1}{HTML}{6F6F6F}
\definecolor{mylightgray2}{HTML}{999999}
\definecolor{mylightgray3}{HTML}{B4B4B4}
\definecolor{mylightgray4}{HTML}{DCDCDC}

\usepackage[hidelinks]{hyperref}

\usepackage{amssymb}
\usepackage{amsfonts}
\usepackage{mathrsfs}
\usepackage{xspace}
\usepackage{bm}
\usepackage{upgreek}

\newcommand{\safemath}[2]{\newcommand{#1}{\ensuremath{#2}\xspace}}

\safemath{\bma}{\mathbf{a}}
\safemath{\bmb}{\mathbf{b}}
\safemath{\bmc}{\mathbf{c}}
\safemath{\bmd}{\mathbf{d}}
\safemath{\bme}{\mathbf{e}}
\safemath{\bmf}{\mathbf{f}}
\safemath{\bmg}{\mathbf{g}}
\safemath{\bmh}{\mathbf{h}}
\safemath{\bmi}{\mathbf{i}}
\safemath{\bmj}{\mathbf{j}}
\safemath{\bmk}{\mathbf{k}}
\safemath{\bml}{\mathbf{l}}
\safemath{\bmm}{\mathbf{m}}
\safemath{\bmn}{\mathbf{n}}
\safemath{\bmo}{\mathbf{o}}
\safemath{\bmp}{\mathbf{p}}
\safemath{\bmq}{\mathbf{q}}
\safemath{\bmr}{\mathbf{r}}
\safemath{\bms}{\mathbf{s}}
\safemath{\bmt}{\mathbf{t}}
\safemath{\bmu}{\mathbf{u}}
\safemath{\bmv}{\mathbf{v}}
\safemath{\bmw}{\mathbf{w}}
\safemath{\bmx}{\mathbf{x}}
\safemath{\bmy}{\mathbf{y}}
\safemath{\bmz}{\mathbf{z}}
\safemath{\bmzero}{\mathbf{0}}
\safemath{\bmone}{\mathbf{1}}

\bmdefine{\biad}{a}
\bmdefine{\bibd}{b}
\bmdefine{\bicd}{c}
\bmdefine{\bidd}{d}
\bmdefine{\bied}{e}
\bmdefine{\bifd}{f}
\bmdefine{\bigd}{g}
\bmdefine{\bihd}{h}
\bmdefine{\biid}{i}
\bmdefine{\bijd}{j}
\bmdefine{\bikd}{k}
\bmdefine{\bild}{l}
\bmdefine{\bimd}{m}
\bmdefine{\bind}{n}
\bmdefine{\biod}{o}
\bmdefine{\bipd}{p}
\bmdefine{\biqd}{q}
\bmdefine{\bird}{r}
\bmdefine{\bisd}{s}
\bmdefine{\bitd}{t}
\bmdefine{\biud}{u}
\bmdefine{\bivd}{v}
\bmdefine{\biwd}{w}
\bmdefine{\bixd}{x}
\bmdefine{\biyd}{y}
\bmdefine{\bizd}{z}

\bmdefine{\bixid}{\xi}
\bmdefine{\bilambdad}{\lambda}
\bmdefine{\bimud}{\mu}
\bmdefine{\bithetad}{\theta}
\bmdefine{\biphid}{\phi}
\bmdefine{\bideltad}{\delta}

\safemath{\bmia}{\biad}
\safemath{\bmib}{\bibd}
\safemath{\bmic}{\bicd}
\safemath{\bmid}{\bidd}
\safemath{\bmie}{\bied}
\safemath{\bmif}{\bifd}
\safemath{\bmig}{\bigd}
\safemath{\bmih}{\bihd}
\safemath{\bmii}{\biid}
\safemath{\bmij}{\bijd}
\safemath{\bmik}{\bikd}
\safemath{\bmil}{\bild}
\safemath{\bmim}{\bimd}
\safemath{\bmin}{\bind}
\safemath{\bmio}{\biod}
\safemath{\bmip}{\bipd}
\safemath{\bmiq}{\biqd}
\safemath{\bmir}{\bird}
\safemath{\bmis}{\bisd}
\safemath{\bmit}{\bitd}
\safemath{\bmiu}{\biud}
\safemath{\bmiv}{\bivd}
\safemath{\bmiw}{\biwd}
\safemath{\bmix}{\bixd}
\safemath{\bmiy}{\biyd}
\safemath{\bmiz}{\bizd}

\safemath{\bmxi}{\bixid}
\safemath{\bmlambda}{\bilambdad}
\safemath{\bmmu}{\bimud}
\safemath{\bmtheta}{\bithetad}
\safemath{\bmphi}{\biphid}
\safemath{\bmdelta}{\bideltad}

\safemath{\bA}{\mathbf{A}}
\safemath{\bB}{\mathbf{B}}
\safemath{\bC}{\mathbf{C}}
\safemath{\bD}{\mathbf{D}}
\safemath{\bE}{\mathbf{E}}
\safemath{\bF}{\mathbf{F}}
\safemath{\bG}{\mathbf{G}}
\safemath{\bH}{\mathbf{H}}
\safemath{\bI}{\mathbf{I}}
\safemath{\bJ}{\mathbf{J}}
\safemath{\bK}{\mathbf{K}}
\safemath{\bL}{\mathbf{L}}
\safemath{\bM}{\mathbf{M}}
\safemath{\bN}{\mathbf{N}}
\safemath{\bO}{\mathbf{O}}
\safemath{\bP}{\mathbf{P}}
\safemath{\bQ}{\mathbf{Q}}
\safemath{\bR}{\mathbf{R}}
\safemath{\bS}{\mathbf{S}}
\safemath{\bT}{\mathbf{T}}
\safemath{\bU}{\mathbf{U}}
\safemath{\bV}{\mathbf{V}}
\safemath{\bW}{\mathbf{W}}
\safemath{\bX}{\mathbf{X}}
\safemath{\bY}{\mathbf{Y}}
\safemath{\bZ}{\mathbf{Z}}

\safemath{\bZero}{\mathbf{0}}
\safemath{\bOne}{\mathbf{1}}
\safemath{\bDelta}{\mathbf{\Delta}}
\safemath{\bLambda}{\mathbf{\UpLambda}}
\safemath{\bPhi}{\mathbf{\Upphi}}
\safemath{\bSigma}{\mathbf{\Upsigma}}
\safemath{\bOmega}{\mathbf{\Upomega}}
\safemath{\bTheta}{\mathbf{\Uptheta}}

\bmdefine{\biAd}{A}
\bmdefine{\biBd}{B}
\bmdefine{\biCd}{C}
\bmdefine{\biDd}{D}
\bmdefine{\biEd}{E}
\bmdefine{\biFd}{F}
\bmdefine{\biGd}{G}
\bmdefine{\biHd}{H}
\bmdefine{\biId}{I}
\bmdefine{\biJd}{J}
\bmdefine{\biKd}{K}
\bmdefine{\biLd}{L}
\bmdefine{\biMd}{M}
\bmdefine{\biOd}{N}
\bmdefine{\biPd}{O}
\bmdefine{\biQd}{P}
\bmdefine{\biRd}{R}
\bmdefine{\biSd}{S}
\bmdefine{\biTd}{T}
\bmdefine{\biUd}{U}
\bmdefine{\biVd}{V}
\bmdefine{\biWd}{W}
\bmdefine{\biXd}{X}
\bmdefine{\biYd}{Y}
\bmdefine{\biZd}{Z}

\bmdefine{\biDelta}{\Delta}
\bmdefine{\biLambda}{\Lambda}
\bmdefine{\biPhi}{\Phi}
\bmdefine{\biSigma}{\Sigma}
\bmdefine{\biOmega}{\Omega}
\bmdefine{\biTheta}{\Theta}

\safemath{\bimA}{\biAd}
\safemath{\bimB}{\biBd}
\safemath{\bimC}{\biCd}
\safemath{\bimD}{\biDd}
\safemath{\bimE}{\biEd}
\safemath{\bimF}{\biFd}
\safemath{\bimG}{\biGd}
\safemath{\bimH}{\biHd}
\safemath{\bimI}{\biId}
\safemath{\bimJ}{\biJd}
\safemath{\bimK}{\biKd}
\safemath{\bimL}{\biLd}
\safemath{\bimM}{\biMd}
\safemath{\bimN}{\biNd}
\safemath{\bimO}{\biOd}
\safemath{\bimP}{\biPd}
\safemath{\bimQ}{\biQd}
\safemath{\bimR}{\biRd}
\safemath{\bimS}{\biSd}
\safemath{\bimT}{\biTd}
\safemath{\bimU}{\biUd}
\safemath{\bimV}{\biVd}
\safemath{\bimW}{\biWd}
\safemath{\bimX}{\biXd}
\safemath{\bimY}{\biYd}
\safemath{\bimZ}{\biZd}

\safemath{\bimDelta}{\biDelta}
\safemath{\bimLambda}{\biLambda}
\safemath{\bimPhi}{\biPhi}
\safemath{\bimSigma}{\biSigma}
\safemath{\bimOmega}{\biOmega}
\safemath{\bimTheta}{\biTheta}

\safemath{\setA}{\mathcal{A}}
\safemath{\setB}{\mathcal{B}}
\safemath{\setC}{\mathcal{C}}
\safemath{\setD}{\mathcal{D}}
\safemath{\setE}{\mathcal{E}}
\safemath{\setF}{\mathcal{F}}
\safemath{\setG}{\mathcal{G}}
\safemath{\setH}{\mathcal{H}}
\safemath{\setI}{\mathcal{I}}
\safemath{\setJ}{\mathcal{J}}
\safemath{\setK}{\mathcal{K}}
\safemath{\setL}{\mathcal{L}}
\safemath{\setM}{\mathcal{M}}
\safemath{\setN}{\mathcal{N}}
\safemath{\setO}{\mathcal{O}}
\safemath{\setP}{\mathcal{P}}
\safemath{\setQ}{\mathcal{Q}}
\safemath{\setR}{\mathcal{R}}
\safemath{\setS}{\mathcal{S}}
\safemath{\setT}{\mathcal{T}}
\safemath{\setU}{\mathcal{U}}
\safemath{\setV}{\mathcal{V}}
\safemath{\setW}{\mathcal{W}}
\safemath{\setX}{\mathcal{X}}
\safemath{\setY}{\mathcal{Y}}
\safemath{\setZ}{\mathcal{Z}}
\safemath{\emptySet}{\varnothing}

\safemath{\colA}{\mathscr{A}}
\safemath{\colB}{\mathscr{B}}
\safemath{\colC}{\mathscr{C}}
\safemath{\colD}{\mathscr{D}}
\safemath{\colE}{\mathscr{E}}
\safemath{\colF}{\mathscr{F}}
\safemath{\colG}{\mathscr{G}}
\safemath{\colH}{\mathscr{H}}
\safemath{\colI}{\mathscr{I}}
\safemath{\colJ}{\mathscr{J}}
\safemath{\colK}{\mathscr{K}}
\safemath{\colL}{\mathscr{L}}
\safemath{\colM}{\mathscr{M}}
\safemath{\colN}{\mathscr{N}}
\safemath{\colO}{\mathscr{O}}
\safemath{\colP}{\mathscr{P}}
\safemath{\colQ}{\mathscr{Q}}
\safemath{\colR}{\mathscr{R}}
\safemath{\colS}{\mathscr{S}}
\safemath{\colT}{\mathscr{T}}
\safemath{\colU}{\mathscr{U}}
\safemath{\colV}{\mathscr{V}}
\safemath{\colW}{\mathscr{W}}
\safemath{\colX}{\mathscr{X}}
\safemath{\colY}{\mathscr{Y}}
\safemath{\colZ}{\mathscr{Z}}

\safemath{\opA}{\mathbb{A}}
\safemath{\opB}{\mathbb{B}}
\safemath{\opC}{\mathbb{C}}
\safemath{\opD}{\mathbb{D}}
\safemath{\opE}{\mathbb{E}}
\safemath{\opF}{\mathbb{F}}
\safemath{\opG}{\mathbb{G}}
\safemath{\opH}{\mathbb{H}}
\safemath{\opI}{\mathbb{I}}
\safemath{\opJ}{\mathbb{J}}
\safemath{\opK}{\mathbb{K}}
\safemath{\opL}{\mathbb{L}}
\safemath{\opM}{\mathbb{M}}
\safemath{\opN}{\mathbb{N}}
\safemath{\opO}{\mathbb{O}}
\safemath{\opP}{\mathbb{P}}
\safemath{\opQ}{\mathbb{Q}}
\safemath{\opR}{\mathbb{R}}
\safemath{\opS}{\mathbb{S}}
\safemath{\opT}{\mathbb{T}}
\safemath{\opU}{\mathbb{U}}
\safemath{\opV}{\mathbb{V}}
\safemath{\opW}{\mathbb{W}}
\safemath{\opX}{\mathbb{X}}
\safemath{\opY}{\mathbb{Y}}
\safemath{\opZ}{\mathbb{Z}}
\safemath{\opZero}{\mathbb{O}}
\safemath{\identityop}{\opI}

\safemath{\veca}{\bma}
\safemath{\vecb}{\bmb}
\safemath{\vecc}{\bmc}
\safemath{\vecd}{\bmd}
\safemath{\vece}{\bme}
\safemath{\vecf}{\bmf}
\safemath{\vecg}{\bmg}
\safemath{\vech}{\bmh}
\safemath{\veci}{\bmi}
\safemath{\vecj}{\bmj}
\safemath{\veck}{\bmk}
\safemath{\vecl}{\bml}
\safemath{\vecm}{\bmm}
\safemath{\vecn}{\bmn}
\safemath{\veco}{\bmo}
\safemath{\vecp}{\bmp}
\safemath{\vecq}{\bmq}
\safemath{\vecr}{\bmr}
\safemath{\vecs}{\bms}
\safemath{\vect}{\bmt}
\safemath{\vecu}{\bmu}
\safemath{\vecv}{\bmv}
\safemath{\vecw}{\bmw}
\safemath{\vecx}{\bmx}
\safemath{\vecy}{\bmy}
\safemath{\vecz}{\bmz}

\safemath{\veczero}{\bmzero}
\safemath{\vecone}{\bmone}
\safemath{\vecxi}{\bmxi}
\safemath{\veclambda}{\bmlambda}
\safemath{\vecmu}{\bmmu}
\safemath{\vectheta}{\bmtheta}
\safemath{\vecphi}{\bmphi}
\safemath{\vecdelta}{\bmdelta}

\safemath{\matA}{\bA}
\safemath{\matB}{\bB}
\safemath{\matC}{\bC}
\safemath{\matD}{\bD}
\safemath{\matE}{\bE}
\safemath{\matF}{\bF}
\safemath{\matG}{\bG}
\safemath{\matH}{\bH}
\safemath{\matI}{\bI}
\safemath{\matJ}{\bJ}
\safemath{\matK}{\bK}
\safemath{\matL}{\bL}
\safemath{\matM}{\bM}
\safemath{\matN}{\bN}
\safemath{\matO}{\bO}
\safemath{\matP}{\bP}
\safemath{\matQ}{\bQ}
\safemath{\matR}{\bR}
\safemath{\matS}{\bS}
\safemath{\matT}{\bT}
\safemath{\matU}{\bU}
\safemath{\matV}{\bV}
\safemath{\matW}{\bW}
\safemath{\matX}{\bX}
\safemath{\matY}{\bY}
\safemath{\matZ}{\bZ}
\safemath{\matzero}{\bmzero}

\safemath{\matDelta}{\bDelta}
\safemath{\matLambda}{\bLambda}
\safemath{\matPhi}{\bPhi}
\safemath{\matSigma}{\bSigma}
\safemath{\matOmega}{\bOmega}
\safemath{\matTheta}{\bTheta}

\safemath{\matidentity}{\matI}
\safemath{\matone}{\matO}

\safemath{\rnda}{A}
\safemath{\rndb}{B}
\safemath{\rndc}{C}
\safemath{\rndd}{D}
\safemath{\rnde}{E}
\safemath{\rndf}{F}
\safemath{\rndg}{G}
\safemath{\rndh}{H}
\safemath{\rndi}{I}
\safemath{\rndj}{J}
\safemath{\rndk}{K}
\safemath{\rndl}{L}
\safemath{\rndm}{M}
\safemath{\rndn}{N}
\safemath{\rndo}{O}
\safemath{\rndp}{P}
\safemath{\rndq}{Q}
\safemath{\rndr}{R}
\safemath{\rnds}{S}
\safemath{\rndt}{T}
\safemath{\rndu}{U}
\safemath{\rndv}{V}
\safemath{\rndw}{W}
\safemath{\rndx}{X}
\safemath{\rndy}{Y}
\safemath{\rndz}{Z}

\safemath{\rveca}{\bimA}
\safemath{\rvecb}{\bimB}
\safemath{\rvecc}{\bimC}
\safemath{\rvecd}{\bimD}
\safemath{\rvece}{\bimE}
\safemath{\rvecf}{\bimF}
\safemath{\rvecg}{\bimG}
\safemath{\rvech}{\bimH}
\safemath{\rveci}{\bimI}
\safemath{\rvecj}{\bimJ}
\safemath{\rveck}{\bimK}
\safemath{\rvecl}{\bimL}
\safemath{\rvecm}{\bimM}
\safemath{\rvecn}{\bimN}
\safemath{\rveco}{\bomO}
\safemath{\rvecp}{\bimP}
\safemath{\rvecq}{\bimQ}
\safemath{\rvecr}{\bimR}
\safemath{\rvecs}{\bimS}
\safemath{\rvect}{\bimT}
\safemath{\rvecu}{\bimU}
\safemath{\rvecv}{\bimV}
\safemath{\rvecw}{\bimW}
\safemath{\rvecx}{\bimX}
\safemath{\rvecy}{\bimY}
\safemath{\rvecz}{\bimZ}

\safemath{\rvecxi}{\bmxi}
\safemath{\rveclambda}{\bmlambda}
\safemath{\rvecmu}{\bmmu}
\safemath{\rvectheta}{\bmtheta}
\safemath{\rvecphi}{\bmphi}

\safemath{\rmatA}{\bimA}
\safemath{\rmatB}{\bimB}
\safemath{\rmatC}{\bimC}
\safemath{\rmatD}{\bimD}
\safemath{\rmatE}{\bimE}
\safemath{\rmatF}{\bimF}
\safemath{\rmatG}{\bimG}
\safemath{\rmatH}{\bimH}
\safemath{\rmatI}{\bimI}
\safemath{\rmatJ}{\bimJ}
\safemath{\rmatK}{\bimK}
\safemath{\rmatL}{\bimL}
\safemath{\rmatM}{\bimM}
\safemath{\rmatN}{\bimN}
\safemath{\rmatO}{\bimO}
\safemath{\rmatP}{\bimP}
\safemath{\rmatQ}{\bimQ}
\safemath{\rmatR}{\bimR}
\safemath{\rmatS}{\bimS}
\safemath{\rmatT}{\bimT}
\safemath{\rmatU}{\bimU}
\safemath{\rmatV}{\bimV}
\safemath{\rmatW}{\bimW}
\safemath{\rmatX}{\bimX}
\safemath{\rmatY}{\bimY}
\safemath{\rmatZ}{\bimZ}

\safemath{\rmatDelta}{\bimDelta}
\safemath{\rmatLambda}{\bimLambda}
\safemath{\rmatPhi}{\bimPhi}
\safemath{\rmatSigma}{\bimSigma}
\safemath{\rmatOmega}{\bimOmega}
\safemath{\rmatTheta}{\bimTheta}

\usepackage{amssymb}
\usepackage{amsfonts}
\usepackage{mathrsfs}
\usepackage{xspace}
\usepackage{bm}
\usepackage{fancyref}
\usepackage{textcomp}

\usepackage{multirow}
\usepackage{stmaryrd}

\newenvironment{textbmatrix}{	\setlength{\arraycolsep}{2.5pt}%
								\big[\begin{matrix}}{\end{matrix}\big]%
								\raisebox{0.08ex}{\vphantom{M}}}

\def\be{\begin{equation}}
\def\ee{\end{equation}}
\def\een{\nonumber \end{equation}}
\def\mat{\begin{bmatrix}}
\def\emat{\end{bmatrix}}
\def\btm{\begin{textbmatrix}}
\def\etm{\end{textbmatrix}}

\def\ba#1\ea{\begin{align}#1\end{align}}
\def\bas#1\eas{\begin{align*}#1\end{align*}}
\def\bs#1\es{\begin{split}#1\end{split}}
\def\bg#1\eg{\begin{gather}#1\end{gather}}
\def\bml#1\eml{\begin{multline}#1\end{multline}}
\def\bi#1\ei{\begin{itemize}#1\end{itemize}}

\newcommand{\lefto}{\mathopen{}\left}

\DeclareMathOperator{\sinc}{sinc}			
\DeclareMathOperator*{\argmin}{arg\;min}		
\DeclareMathOperator{\kron}{\otimes}			
\DeclareMathOperator{\Exop}{\opE}			

\newcommand{\Ex}[2]{\ensuremath{\Exop_{#1}\lefto[#2\right]}} 	

\newcommand{\tp}[1]{\ensuremath{#1^{\text{T}}}} 		
\newcommand{\herm}[1]{\ensuremath{#1^{\text{H}}}} 	

\safemath{\dirac}{\delta}					
\safemath{\krond}{\dirac}					

\safemath{\upto}{\uparrow}
\safemath{\downto}{\downarrow}
\safemath{\iu}{j}							
\safemath{\ev}{\lambda}						
\safemath{\hilseqspace}{l^{2}}				
\newcommand{\banachfunspace}[1]{\setL^{#1}}	
\safemath{\hilfunspace}{\banachfunspace{2}}	

\safemath{\SNR}{\textit{SNR}} 				
\safemath{\PAR}{\textit{PAR}} 				
\safemath{\No}{N_0}							
\safemath{\Es}{E_s}							
\safemath{\Eb}{E_b}							
\safemath{\EbNo}{\frac{\Eb}{\No}}
\safemath{\EsNo}{\frac{\Es}{\No}}

\DeclareMathOperator{\CHop}{\ensuremath{\opH}} 
\safemath{\tvir}{\rndh_{\CHop}}				
\safemath{\tvtf}{\rndl_{\CHop}}				
\safemath{\spf}{\rnds_{\CHop}}				
\safemath{\bff}{H_{\CHop}}					

\safemath{\ircf}{r_{h}}						
\safemath{\tftvcf}{r_{s}}					
\safemath{\tfcf}{r_{l}}						
\safemath{\bfcf}{r_{H}}						

\safemath{\tcorr}{c_h}						
\safemath{\scf}{c_{s}}						
\safemath{\tfcorr}{c_{l}}					
\safemath{\fcorr}{c_{H}}						

\safemath{\mi}{I}							
\safemath{\capacity}{C}						

\safemath{\normal}{\mathcal{N}}			
\safemath{\jpg}{\mathcal{CN}}			
\safemath{\mchain}{\leftrightarrow}		

\safemath{\dB}{\,\mathrm{dB}}
\safemath{\dBm}{\,\mathrm{dBm}}
\safemath{\Hz}{\,\mathrm{Hz}}
\safemath{\kHz}{\,\mathrm{kHz}}
\safemath{\MHz}{\,\mathrm{MHz}}
\safemath{\GHz}{\,\mathrm{GHz}}
\safemath{\s}{\,\mathrm{s}}
\safemath{\ms}{\,\mathrm{ms}}
\safemath{\mus}{\,\mathrm{\text{\textmu}s}}
\safemath{\ns}{\,\mathrm{ns}}
\safemath{\ps}{\,\mathrm{ps}}
\safemath{\meter}{\,\mathrm{m}}
\safemath{\mm}{\,\mathrm{mm}}
\safemath{\cm}{\,\mathrm{cm}}
\safemath{\m}{\,\mathrm{m}}
\safemath{\W}{\,\mathrm{W}}
\safemath{\mW}{\, \mathrm{mW}}
\safemath{\J}{\,\mathrm{J}}
\safemath{\K}{\,\mathrm{K}}
\safemath{\bit}{\,\mathrm{bit}}
\safemath{\nat}{\,\mathrm{nat}}

\safemath{\define}{\triangleq}			

\safemath{\equivalent}{\sim}
\safemath{\distas}{\sim}					
\safemath{\sdiff}{\Delta}				

\safemath{\reals}{\mathbb{R}}
\safemath{\positivereals}{\reals_{+}}
\safemath{\integers}{\mathbb{Z}}
\safemath{\posint}{\integers_{+}}
\safemath{\naturals}{\mathbb{N}}
\safemath{\posnaturals}{\naturals_{+}}
\safemath{\complexset}{\mathbb{C}}
\safemath{\rationals}{\mathbb{Q}}

\newcommand*{\fancyrefapplabelprefix}{app}		
\newcommand*{\fancyrefthmlabelprefix}{thm}		
\newcommand*{\fancyreflemlabelprefix}{lem}		
\newcommand*{\fancyrefcorlabelprefix}{cor}		
\newcommand*{\fancyrefdeflabelprefix}{def}		
\newcommand*{\fancyrefproplabelprefix}{prop}		
\newcommand*{\fancyrefexmpllabelprefix}{exmpl}
\newcommand*{\fancyrefalglabelprefix}{alg}		
\newcommand*{\fancyreftbllabelprefix}{tbl}		

\frefformat{vario}{\fancyrefseclabelprefix}{Section~#1}
\frefformat{vario}{\fancyrefthmlabelprefix}{Theorem~#1}
\frefformat{vario}{\fancyreftbllabelprefix}{Table~#1}
\frefformat{vario}{\fancyreflemlabelprefix}{Lemma~#1}
\frefformat{vario}{\fancyrefcorlabelprefix}{Corollary~#1}
\frefformat{vario}{\fancyrefdeflabelprefix}{Definition~#1}
\frefformat{vario}{\fancyreffiglabelprefix}{Fig.~#1}
\frefformat{vario}{\fancyrefapplabelprefix}{Appendix~#1}
\frefformat{vario}{\fancyrefeqlabelprefix}{(#1)}
\frefformat{vario}{\fancyrefproplabelprefix}{Proposition~#1}
\frefformat{vario}{\fancyrefexmpllabelprefix}{Example~#1}
\frefformat{vario}{\fancyrefalglabelprefix}{Algorithm~#1}

 \newtheorem{thm}{Theorem}

 \newtheorem{lem}[thm]{Lemma}
 
 \newtheorem{remark}{Remark}
 \newtheorem*{remark*}{Remark}

\safemath{\dictab}{[\,\dicta\,\,\dictb\,]}

\safemath{\ysig}{\bmy}
\safemath{\ysighat}{\hat{\ysig}}
\safemath{\ysigdim}{M}
\safemath{\xsig}{\bmx}
\safemath{\xsigdim}{N}
\safemath{\nx}{n_x}
\safemath{\zsig}{\bmz}
\safemath{\zsigdim}{\ysigdim}
\safemath{\rsig}{\bmr}
\safemath{\Adict}{\bA}
\safemath{\Adicttilde}{\widetilde{\Adict}}
\safemath{\Adictdim}{\outputdim\times\xsigdim}
\safemath{\avec}{\bma}
\safemath{\avectilde}{\tilde{\avec}}
\safemath{\Bdict}{\bB}
\safemath{\Bdicttilde}{\widetilde{\Bdict}}
\safemath{\Cdict}{\bC}
\safemath{\cvec}{\bmc}
\safemath{\Ddict}{\bD}
\safemath{\Ddictdim}{\ysigdim\times\xsigdim}
\safemath{\dvec}{\bmd}
\safemath{\Ddicttilde}{\widetilde{\bD}}
\safemath{\Bonb}{\bB}
\safemath{\bvec}{\bmb}
\safemath{\Bonbdim}{\ysigdim\times\ysigdim}
\safemath{\noise}{\bmn}
\safemath{\noisedim}{\ysigim}
\safemath{\err}{\bme}
\safemath{\errdim}{\ysigdim}
\safemath{\errset}{\setE}
\safemath{\nerr}{n_e}
\safemath{\delop}{\bP_\errset}
\safemath{\delopc}{\bP_{{\errset}^c}}

\safemath{\cplxi}{\imath}
\safemath{\cplxj}{\jmath}

\safemath{\dict}{\matD}
\safemath{\inputdim}{N}		
\safemath{\outputdim}{M}		
\safemath{\sparsity}{S}	
\safemath{\inputdimA}{{N_a}}	
\safemath{\inputdimB}{{N_b}}	
\safemath{\elemA}{{n_a}}	
\safemath{\elemB}{{n_b}}	
\safemath{\resA}{\matR_a}	
\safemath{\resB}{\matR_b}	
\safemath{\subD}{\matS} 
\safemath{\subA}{\matS_a} 
\safemath{\subB}{\matS_b} 
\safemath{\dicta}{\matA} 	
\safemath{\dictb}{\matB} 	
\safemath{\hollowS}{H}
\safemath{\hollowA}{H_a}
\safemath{\hollowB}{H_b}
\safemath{\cross}{Z}
\safemath{\coh}{\mu_d}			
\safemath{\coha}{\mu_a}			
\safemath{\cohb}{\mu_b}			
\safemath{\mubs}{\nu}	
\safemath{\cohm}{\mu_m} 
\safemath{\dictset}{\setD}	
\safemath{\dictsetp}{\dictset(\coh,\coha,\cohb)}	
\safemath{\dictsetgen}{\dictset_\text{gen}}
\safemath{\dictsetgenp}{\dictsetgen(\coh)}
\safemath{\dictsetonb}{\dictset_\text{onb}}
\safemath{\dictsetonbp}{\dictsetonb(\coh)}

\safemath{\leftside}{U}
\safemath{\rightsideA}{R_a}
\safemath{\rightsideB}{R_b}

\safemath{\indexS}{\setI_S} 

\safemath{\na}{n_a}			
\safemath{\nb}{n_b}			
\safemath{\coeffa}{p_i}	
\safemath{\coeffb}{q_j}	
\safemath{\seta}{\setP}		
\safemath{\setb}{\setQ}     
\safemath{\setw}{\setW}	
\safemath{\setz}{\setZ}	
\safemath{\cola}{\veca}		
\safemath{\colb}{\vecb}		
\safemath{\cold}{\vecd}		
\safemath{\inputvec}{\vecx} 	
\safemath{\error}{\vece}	
\safemath{\noiseout}{\vecz} 	
\safemath{\inputvecel}{x}
\safemath{\inputveca}{\vecx_a}
\safemath{\inputvecb}{\vecx_b}
\safemath{\outputvec}{\vecy}	
\safemath{\lambdamin}{\lambda_{\mathrm{min}}}

\safemath{\elltwo}{\ell_2}
\safemath{\ellone}{\ell_1}
\safemath{\ellzero}{\ell_0}
\safemath{\ellinf}{\ell_\infty}
\safemath{\ellinftilde}{\ell_{\widetilde\infty}}
\safemath{\licard}{Z(\coh,\coha,\cohb)}
\safemath{\xsol}{\hat{x}}
\safemath{\xbord}{x_b}		
\safemath{\xstat}{x_s}		
\safemath{\xstatLone}{\tilde{x}_s}
\safemath{\order}{\mathcal{O}} 
\safemath{\scales}{\Theta} 
\safemath{\ones}{\mathbf{1}} 
\safemath{\zeroes}{\mathbf{0}} 
\safemath{\thlone}{\kappa(\coh,\cohb)} 
\safemath{\constoneA}{\delta} 
\safemath{\constoneB}{\epsilon} 
\safemath{\nlarge}{L}				   
\safemath{\sumlarge}{S_\nlarge}
\safemath{\maxlarger}{P_\nlarge}	   
\safemath{\Pzero}{\textrm{P0}}	
\safemath{\Pone}{\textrm{P1}}
\safemath{\vecfir}{\vecw}			 
\safemath{\vecsec}{\vecz}
\safemath{\elvecfir}{w}              
\safemath{\elvecsec}{z}				 
\safemath{\nlargefir}{n}
\safemath{\normout}{\gamma}
\safemath{\auxfun}{h}
\safemath{\supp}{\textrm{supp}}

\safemath{\indexa}{\ell}
\safemath{\indexb}{r}
\safemath{\indexc}{i}
\safemath{\indexd}{j}

\safemath{\project}{P}

\usepackage{framed}

\IEEEoverridecommandlockouts
\allowdisplaybreaks 

\safemath{\Nat}{{N}}
\safemath{\Nrow}{{N_\text{row}}}
\safemath{\Ncol}{{N_\text{col}}}
\safemath{\Npil}{{N_\text{pil}}}
\safemath{\Nsc}{{N_\text{sc}}}
\safemath{\Fpil}{{\setF_\text{pil}}}
\safemath{\btheta}{\boldsymbol{\theta}}
\safemath{\bnu}{{\boldsymbol{\nu}}}
\renewcommand{\bSigma}{{\boldsymbol{\Sigma}}}
\renewcommand{\bLambda}{{\boldsymbol{\Lambda}}}
\renewcommand{\No}{{\mathsf{N}_0}}

\safemath{\bsfa}{\boldsymbol{\mathsf{a}}}
\safemath{\bsfS}{\boldsymbol{\mathsf{S}}}
\newcommand{\tinysquare}{\scriptscriptstyle{\square}}

\begin{document}
\bstctlcite{IEEEexample:BSTcontrol} 

\title{Sparse MIMO-OFDM Channel Estimation\\ via RKHS Regularization}

\author{
\IEEEauthorblockN{James Delfeld$^{\sharp}$, Gian Marti$^{\flat,\natural}$, Chris Dick$^{\sharp}$}\\
\IEEEauthorblockA{$^{\sharp}$NVIDIA, $^{\flat}$ETH Zurich}
\thanks{$^{\natural}$Work done during an internship at NVIDIA.}
\thanks{Discussions with F. A. Aoudia, M. Belgiovine, S. Cammerer, J. Hoydis, L.~Maggi, and C.~Studer are gratefully acknowledged.}
}
\maketitle
\thispagestyle{empty}

\begin{abstract}
    We propose a method for channel estimation in  multiple-input multiple-output (MIMO) orthogonal frequency-division 
    multiplexing (OFDM) wireless communication systems. 
    The method exploits the band-sparsity of wireless channels in the delay-beamspace domain by solving a regularized optimization 
    problem in a reproducing kernel Hilbert space (RKHS). A~suitable representer theorem allows us to
    transform the infinite-dimensional optimization problem into a finite-dimensional one, which we then approximate with a
    low-dimensional surrogate. We solve the resulting optimization problem using a forward-backward splitting (FBS)-based algorithm. 
    By exploiting the problem's modulation structure, we achieve a 
    computational complexity per iteration that is quasi-linear in the number of unknown variables.
    We also propose a data-driven deep-unfolding based extension to improve the
    performance at a reduced number of iterations. 
    We evaluate our channel estimators on ray-traced channels generated with SionnaRT. 
    The results show that our methods significantly outperform linear methods such as linear minimum mean squared error (LMMSE) 
    channel estimation based on aggregate channel statistics, both in terms of raw estimation accuracy as well as in downstream performance.
\end{abstract}

\section{Introduction}

Obtaining accurate channel estimates is~of paramount importance for many wireless communication systems~\cite{medard2000effect, lapidoth2002fading}. 
MIMO-OFDM systems estimate the channel on the basis of noisy measurements taken at different frequencies 
(corresponding to the different subcarriers) and spatial positions (corresponding to the different receive antennas)~\cite{larsen2009performance, ozdemir2007channel}.
Estimation accuracy can be improved by leveraging basic characteristics of wireless channels, such as 
correlation across time, frequency, and space \cite{ozdemir2007channel, van1995channel}, 
or sparsity in a suitable representation \cite{cotter2002sparse, berger2010application}. 
To leverage these characteristics, one can use classical model-based approaches \cite{chang2002model, xie2016overview}, data-driven (often black-box) approaches \cite{le2021deep, soltani2019deep}, 
or hybrid approaches that combine certain aspects of the other two \cite{neumann2018learning}.

\subsection{Contributions}
We propose a model-based channel estimator for MIMO-OFDM that leverages the band-sparsity of wireless channels in the delay-beamspace domain. To this end, our method solves an optimization problem with a convex band-sparsity promoting regularizer
in an RKHS. Using a suitable representer theorem, the infinite-dimensional 
optimization problem is transformed into an equivalent finite- (but \mbox{high-)dimensional} problem, which we then approximate with a 
low-dimensional surrogate. To solve this resulting optimization problem, we propose an efficient FBS-based \cite{goldstein16a} 
algorithm that exploits the problem's modulation structure and achieves a computational complexity that is quasi-linear
in the number of unknown variables (i.e., $O(n\log n)$, where $n$ is the number of unknown variables).
To remove the bias introduced by convex sparsifying regularization, we use a subsequent debiasing stage that performs Tikhonov-regularized
channel estimation while constraining the estimate to the previously detected support. 

We then present a data-driven extension of our RKHS channel estimator based on deep unfolding \cite{balatsoukas2019deep}. 
In this extension, the convex regularizer is replaced with a non-convex regularizer with additional hyperparameters that are learned 
from data. Compared to the vanilla RKHS channel estimator, the data-driven 
extension achieves improved performance in fewer iterations (i.e., at lower computational
complexity).\footnote{Among other things, the non-convex regularizer enables dispensing with the debiasing stage entirely.}

We evaluate our algorithms in simulations on wireless channels generated with SionnaRT~\cite{aoudia2025sionna}. 
Our results show that our methods outperform least squares (LS) channel estimation by 26\,dB and 27\,dB, respectively, 
and empirical LMMSE channel estimation based on aggregate channel statistics by 15\,dB and 16\,dB, respectively, 
in the signal-to-noise ratio~(SNR) required to achieve a normalized mean squared error (NMSE) estimation accuracy of 1\%.
Simulations also show that these gains translate to increased achievable rates in the downlink when the obtained channel estimates
are used for maximum ratio combining (MRC) in downlink precoding.

\subsection{Related Work}
Much effort has been devoted to the development of high-performance channel estimators. 
In model-based approaches, sparsity has established itself as a central theme besides the
reliance on temporal, spectral, and spatial correlations \cite{ozdemir2007channel, van1995channel, sionna_chest}.
Sparsity can be leveraged in various forms, such as in a 
low-rank assumption on the channel \cite{xie2016overview, eliasi2017low, li2017millimeter},
in an explicitly sparse representation such as the beamspace (or angular) domain \cite{lee2016channel, wang2017sparse, mirfarshbafan19a, wen2014channel}, 
or in the form of an underlying sparse generative model, e.g., Gaussian mixtures \cite{wen2014channel, bock2025physics, bock2025sparse}.
We emphasize that these different forms are not mutually exclusive, but rather provide different viewpoints on common underlying
characteristics, and in fact are often equivalent. 
In contrast to these methods, which rely on discrete notions of sparsity, our method uses a continuous notion 
of band-sparsity that better reflects the continuous nature of electromagnetic wave propagation.

Besides model-based methods, data-driven methods seem to promise superior performance at the cost 
of training-data dependency as well as, typically, increased computational complexity. 
Suggested approaches include fully connected deep neural networks \cite{le2021deep, melgar2022deep, ge2021deep},  
convolutional neural networks inspired by computer vision \cite{soltani2019deep}, 
variational auto-encoders~\cite{baur2022variational}, 
or neural networks used as regularizers \cite{balevi2020massive}.

There is no hard boundary between model-based and data-driven approaches, and proposed methods often straddle the two 
in order to reduce the dependence on training data and/or the computational complexity while maintaining the performance 
of data-driven methods.
Examples of such hybrid methods include Gaussian mixture-based methods \cite{koller2022asymptotically, fesl2022channel} 
as~well as the deep unfolding-based dictionary learning methods \cite{yassine2022mpnet, chatelier2023efficient}.
However, the methods in those references rely on training data generated from a statistical channel model that mirrors the 
model of the channel inference process. In contrast, our data-driven extension uses training data from the wireless ray-tracer
SionnaRT \cite{aoudia2025sionna}, so that there is no analytic connection between channel generation and channel inference.

\subsection{Notation}
Matrices are denoted as boldface uppercase letters. Column vectors and tuples are denoted as boldface lowercase letters. 
For a matrix $\bA$, the conjugate is $\bA^\ast$, the transpose is $\tp{\bA}$ and the conjugate transpose is $\herm{\bA}$.
The Kronecker product is $\kron$ and the Hadamard product is $\odot$.
A general norm is denoted~$\|\cdot\|$ while the Euclidean norm of a vector is denoted $\|\bma\|_2$.
The set of integers from $N$ through~$N'$ is denoted $[N:N']$.
The $N\times1$ all-ones vector is denoted $\mathds{1}_N$.

\section{Problem Statement}
\label{sec:problem}
We consider a cellular MIMO-OFDM system where a multi-antenna basestation (BS) estimates
the wireless channel of a single-antenna user equipment (UE) from noisy measurements.

Specifically, we consider a BS with a two-dimensional antenna array consisting of $\Nrow$ rows and $\Ncol$ columns 
(i.e., the total number of BS antennas is $\Nrow \Ncol$), and an OFDM system with subcarrier frequencies 
$\setF$ of which the subset $\Fpil=\big\{f_{m_1} \,|\, m_1\in[1:\Npil]\big\}\subset\setF$ carries pilots. 
The set of positions $\setX$ of the BS's antennas is parametrized as 
\begin{align}
    \!\setX \!=\! \big\{\bmr+ x^{(\text{r})}_{m_2}\, \bme_\text{row} + x^{(\text{c})}_{m_3}\, \bme_\text{col} \,|\, 
    m_2\!\in\![1\!:\!\Nrow], m_3\!\in\![1\!:\!\Ncol] \big\},\! \label{eq:antenna_positions}
\end{align}
where $\bmr\in\opR^3$ is the BS's position in global coordinates, 
where $\bme_\text{row}\in\mathbb{R}^3$ and $\bme_\text{col}\in\mathbb{R}^3$ are unit vectors that determine 
the orientation of the BS's antenna array, 
and where $x^{(\text{r})}_{m_2}$ and $x^{(\text{c})}_{m_3}$ specify the antenna positions 
within the BS-centered two-dimensional coordinate system with basis vectors $\bme_\text{row}$ and $\bme_\text{col}$. 
The BS's goal is to estimate the channel $h(f, x^{(\text{r})}_{m_2}, x^{(\text{c})}_{m_3})$
at all antenna positions and all subcarrier frequencies $f\in\setF$,
where this estimate is formed on the basis of noisy measurements 
\begin{align}
\!\!y(f_{m_1}, x^{(\text{r})}_{m_2}, x^{(\text{c})}_{m_3}) = h(f_{m_1}, x^{(\text{r})}_{m_2}, x^{(\text{c})}_{m_3}) + n(f_{m_1}, x^{(\text{r})}_{m_2}, x^{(\text{c})}_{m_3})\!
\label{eq:measurements_1}
\end{align}
acquired at all antenna positions \emph{but only} at the pilot-carrying subcarrier frequencies $f_{m_1}\in\Fpil$. 
In other words, frequency-domain interpolation is an integral aspect of the considered channel estimation problem.\footnote{In 
this paper, we ignore the dynamic aspects of wireless channels such as Doppler shifts. 
We therefore assume the wireless environment to be static. However, incorporating Doppler shifts into the model would be possible.}
The noise $n(\cdot,\cdot,\cdot)$ in \eqref{eq:measurements_1} is assumed to be i.i.d. circularly-symmetric complex Gaussian 
with variance $\No$.
For notational simplicity, it will sometimes be convenient to write \eqref{eq:measurements_1} equivalently as
\begin{align}
    y(\bmz_\bmm) = h(\bmz_\bmm) + n(\bmz_\bmm), \quad \bmm\in\setM \label{eq:measurements_compact}
\end{align}
where $\bmm\triangleq(m_1,m_2,m_3)$, $\setM\triangleq[1\!:\!\Npil]\times[1\!:\!\Nrow]\times[1\!:\!\Ncol]$, and $\bmz_\bmm\triangleq(f_{m_1}, x^{(\text{r})}_{m_2}, x^{(\text{c})}_{m_3})$.

\section{Band-Sparse Wireless Channel Model} 
\label{sec:model}

\begin{figure}[tp]
\centering
\includegraphics[width=0.8\columnwidth]{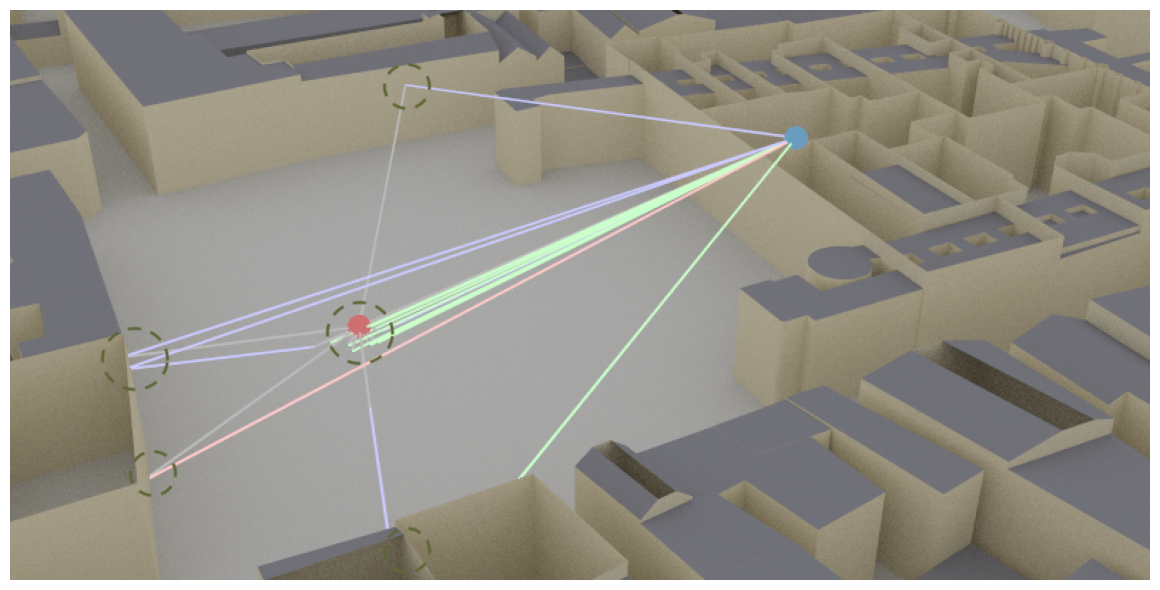}
\caption{Paths between transmitter (shown in red) and receiver (shown in blue). 
The dashed circles highlight clusters of paths with similar angles of arrival and delays.}
\label{fig:paths}
\end{figure}

A radio signal emitted from a transmit position $\bmt\in\mathbb{R}^3$ (identified with the UE position)
typically reaches a receive position $\bmr \in \mathbb{R}^3$ (identified with the BS position)
via a direct path---provided that this path is not obstructed---as well as multiple indirect paths; see \fref{fig:paths}. 
Thus, a transmit signal $s(t)$ leads to a receive signal 
\begin{align}
    y(t) = \sum_{\ell=1}^L \alpha_\ell\, s(t-\tau_\ell) + n(t), \label{eq:equation_1} 
\end{align}
where $L$ is the number of paths, $\alpha_\ell\in\opC$ and $\tau_\ell\in\mathbb{R}$ are the complex path gain and and path delay 
of the $\ell$th path, and $n(t)$ is noise.
Equation \eqref{eq:equation_1} can be rewritten as 
\begin{align}
    y(t) = (\underline{h} \ast s)(t) + n(t),
\end{align}
where 
\begin{align}
    \underline{h}(\tau) = \sum_{\ell=1}^L \alpha_\ell\, \delta(\tau-\tau_\ell) 
\end{align}
is the wireless channel from $\bmt$ to $\bmr$ in the impulse-response representation. 
By taking the Fourier transform, we obtain the channel in the frequency-response representation
\begin{align}
    h(f) = \mathbb{F}_{\tau\to f}\{\underline{h}\}(f) = \sum_{\ell=1}^L \alpha_\ell\, e^{-i2\pi f \tau_\ell}.
\end{align}
Note that $\alpha_\ell$ and $\tau_\ell$ are functions of $\bmt$ and $\bmr$ (as well as of the surrounding environment), 
i.e., we have $\alpha_\ell=\alpha_\ell(\bmt, \bmr)$ and $\tau_\ell=\tau_\ell(\bmt, \bmr)$.
Consider how the channel behaves in the immediate vicinity $\bmr+\Delta\bmr$ of $\bmr$: 
We assume that the paths themselves, as well as their gains $\alpha_\ell$ do not change between $\bmr$ and $\bmr+\Delta\bmr$. 
For the path delays, we obtain $\tau_\ell(\bmt, \bmr+\Delta\bmr) = \tau_\ell(\bmt, \bmr) +  \langle\bme_\bnu(\btheta_\ell), \Delta\bmr\rangle/c$, 
where $c$ is the speed of light and $\bme_\bnu(\btheta_\ell)\in\mathbb{R}^3$ is the unit vector expressing
the direction of wave propagation as a function of the angle of arrival $\btheta_\ell\in [0, 2\pi)\times[0,\pi)$.
We can thus write the channel as a function of the frequency $f$ and the receive position $\bmr+\Delta\bmr$, 
\begin{align}
    h(f,\bmr+\Delta\bmr) 
    &= \sum_{\ell=1}^L \alpha_\ell\, e^{-i2\pi f \tau_\ell}\, e^{-i2\pi \langle\bnu(\btheta_\ell), \Delta\bmr\rangle}, 
    \label{eq:h_freq_space}
\end{align}
where $\bnu(\cdot)=(f/c)\,\bme_\bnu(\cdot)$ is the wave vector. 
In particular, by letting $\Delta\bmr=x^{(\text{r})} \bme_\text{row} + x^{(\text{c})}\bme_\text{col}$, 
where $\bme_\text{row}$ and $\bme_\text{col}$ are the antenna array orientation vectors from \fref{sec:problem},
we obtain the channel in the plane that contains the receive antenna array: 
\begin{align}
&\!\!\! h(f,x^{(\text{r})}, x^{(\text{c})}) \\
&\!\!\! \triangleq h(f,\bmr+x^{(\text{r})} \bme_\text{row} + x^{(\text{c})}\bme_\text{col}) \label{eq:sample_eq} \\
&\!\!\!= \sum_{\ell=1}^L \alpha_\ell\, e^{-i2\pi f \tau_\ell}\, e^{-i2\pi x^{(\text{r})} \langle\bnu(\btheta_\ell), \bme_\text{row}\rangle}
    \,e^{-i2\pi x^{(\text{c})}\langle\bnu(\btheta_\ell), \bme_\text{col}\rangle}\!.\!\! \label{eq:continuous_channel}
\end{align}
Having defined $h(f,x^{(\text{r})}, x^{(\text{c})})= h(f,\bmr+x^{(\text{r})} \bme_\text{row} + x^{(\text{c})}\bme_\text{col})$ 
in \eqref{eq:sample_eq}, we note that the noisy measurements from \eqref{eq:measurements_1} correspond to noisy samples 
of the function in \eqref{eq:continuous_channel} at frequencies $\Fpil$ and positions $\setX$. 
We now state four key assumptions/idealizations: 
\begin{enumerate}
    \item We assume that the transmitting UE, as well as any scatterers, are in the far-field of the BS antenna array.  
    \item We assume that the path gains $\alpha_\ell$ are constant on $\Fpil\times\setX$. 
    \item We assume that the path delays $\tau_\ell$ are constant on $\Fpil$.
    \item We assume that the wavelength $f/c$ is constant on  $\Fpil$.
\end{enumerate}
Under these assumptions, the quantities $\alpha_\ell$, $\tau_\ell$, and $\bnu(\btheta_\ell)$ in \eqref{eq:continuous_channel}
are constants over the domain of interest. 
Let now
\begin{align}
    \underline{h}(\tau, \phi^{(\text{h})}, \phi^{(\text{v})})
    = \mathbb{F}^{-1}_{f\to\tau} \, \mathbb{F}^{-1}_{x^{(\text{r})}\to\phi^{(\text{h})}} \, \mathbb{F}^{-1}_{x^{(\text{c})}\to\phi^{(\text{v})}} 
    \{h\}(f, x^{(\text{r})}, x^{(\text{c})})
\end{align}
be the channel from \eqref{eq:continuous_channel} in delay-beamspace domain.
Then the following holds: 

\begin{mdframed}[style=csstyle]
The delay-beamspace channel $\underline{h}(\tau, \phi^{(\text{h})}, \phi^{(\text{v})})$ is \mbox{sparse} in all three arguments.
\end{mdframed}

Specifically, $\underline{h}(\tau, \phi^{(\text{h})}, \phi^{(\text{v})})=0$ for \mbox{$\tau\notin\{\tau_\ell \,|\, \ell\in[1:L]\}$,}
for $\phi^{(\text{h})}\notin\{\langle\bnu(\btheta_\ell), \bme_\text{row}\rangle \,|\, \ell\in[1:L]\}$, 
as well as for $\phi^{(\text{v})}\notin\{\langle\bnu(\btheta_\ell), \bme_\text{col}\rangle \,|\, \ell\in[1:L]\}$. 
Moreover, paths often appear in clusters with similar delays and angles of arrival, see \fref{fig:paths}.
The total number of such clusters is significantly smaller than the number of paths $L$. 
Thus, if $\frac{T}{2} \geq \max\{|\tau_\ell| \,|\, \ell\in[1\!:\!L]\}$,
$\frac{\Phi^{(\text{h})}}{2} \!\geq\!  \max\{|\langle\bnu(\btheta_\ell), \bme_\text{col}\rangle| \,|\, \ell\in\![1\!:\!L]\}$,
and $\frac{\Phi^{(\text{v})}}{2} \geq  \max\{|\langle\bnu(\btheta_\ell), \bme_\text{row}\rangle| \,|\, \ell\in[1\!:\!L]\}$
are bounds on the possible delays and beams, then we can partition the domain 
$[-\frac{T}{2}, \frac{T}{2}]\times[-\frac{\Phi^{(\text{h})}}{2}, \frac{\Phi^{(\text{h})}}{2}]\times[-\frac{\Phi^{(\text{v})}}{2},\frac{\Phi^{(\text{v})}}{2}]$
of possible delays and beams into boxes
\begin{align}
    \setB_{\bmn} &= \setT_{n_1} \times \setH_{n_2} \times \setV_{n_3},
\end{align}
for $\bmn=(n_1,n_2,n_3)\in\setN=[1\!:\!N_1]\times[1\!:\!N_2]\times[1\!:\!N_3]$, where
\begin{align}
\setT_{n_1} &= \bigg[\frac{(n-1-\frac{N_1}{2})T}{N_1},  \frac{(n-\frac{N_1}{2})T}{N_1} \bigg] \label{eq:T_interval} \\
\setH_{n_2} &= \bigg[\frac{(n-1-\frac{N_2}{2})\Phi^{(\text{h})}}{N_2}, \frac{(n-\frac{N_2}{2})\Phi^{(\text{h})}}{N_2} \bigg] \\
\setV_{n_3} &= \bigg[ \frac{(n-1-\frac{N_3}{2})\Phi^{(\text{v})}}{N_3}, \frac{(n-\frac{N_3}{2})\Phi^{(\text{v})}}{N_3} \bigg].
\end{align}
Thus, we can decompose the channel in the delay-beamspace domain into atoms $h_\bmn, \bmn\in\setN$, as
\begin{align}
    \underline{h}(\tau, \phi^{(\text{h})}, \phi^{(\text{v})})
    &= \sum_{\bmn\in\setN} \underline{h}_{\bmn}(\tau, \phi^{(\text{h})}, \phi^{(\text{v})}),  \label{eq:sparse_decomposition}
\end{align}
where $\textit{supp}(\underline{h}_{n_1,n_2,n_3})\subseteq \setB_{n_1,n_2,n_3}$, 
or in the frequency-space domain as 
\begin{align}
    h(f, x^{(\text{r})}, x^{(\text{c})})
    &= \sum_{\bmn\in\setN} h_{\bmn}(f, x^{(\text{r})}, x^{(\text{c})}), 
\end{align}
where 
\begin{align}
    &h_{\bmn}(f, x^{(\text{r})}, x^{(\text{c})}) \nonumber \\
    &= \mathbb{F}_{\tau\to f} \, \mathbb{F}_{\phi^{(\text{h})}\to x^{(\text{r})}} \, \mathbb{F}_{\phi^{(\text{v})}\to x^{(\text{c})}} 
    \{\underline{h}_{\bmn}\}(\tau, \phi^{(\text{h})}, \phi^{(\text{v})}). \label{eq:h_atoms}
\end{align}
Since the support of $\underline{h}(\tau, \phi^{(\text{h})}, \phi^{(\text{v})})$ in the delay-beamspace domain
tends to be confined to a small number of boxes $\setB_{\bmn}$, most of the atoms in \eqref{eq:sparse_decomposition} will be zero. 
However, the number of supporting boxes is typically not known in advance (just as the number of paths $L$ is typically not known in advance).
Note that, at Nyquist sampling, $N_1$ corresponds to the number $\Npil$ of pilot-carrying subcarriers, 
$N_2$ to the number $\Ncol$ of antenna columns, and $N_3$ to the number $\Nrow$ of antenna rows. 
However, it is also possible to use over- or subsampling.

\section{Sparse Channel Estimation\\via RKHS Regularization}
\label{sec:chest}
Given the modeling assumptions from \fref{sec:model}, it would be desirable to estimate the channel by solving
the regularized optimization problem
\begin{align}
    \mathop{\textnormal{minimize}}_{\substack{\tilde{h}=\sum_{\bmn\in\setN}\tilde{h}_\bmn:\\ \textit{supp}(\underline{\tilde{h}}_\bmn)\in\setB_\bmn}}
    \sum_{\bmm\in\setM} \big| y(\bmz_\bmm) - \tilde{h}(\bmz_\bmm) \big|^2 + \sum_{\bmn\in\setN} \lambda_{\bmn}\mathbf{1}_{\opR_{>0}}(\|\tilde{h}_\bmn\|),
    \label{eq:continuous_problem_1}
\end{align}
where $\mathbf{1}_{\setA}(a)$ is one if $a\in\setA$ and zero otherwise, 
and where the $\lambda_{\bmn}>0$ are regularization coefficients that control the tradeoff 
between the goodness of the fit (captured by the first term in \eqref{eq:continuous_problem_1})
and the band-sparsity (captured by the second term in \eqref{eq:continuous_problem_1}) of the obtained channel estimate.\footnote{
\label{footnote:insinuate}
It seems natural to use the same coefficient $\lambda_{\bmn}$ for all $\bmn\in\setN$, but we will see the advantages
of not doing so in \fref{sec:deep_unfolding}.
}
Note that, by Parseval's identity, the norm $\|\tilde{h}_\bmn\|$ in \eqref{eq:continuous_problem_1} is equal 
to~$\|\underline{\tilde{h}}_\bmn\|$. Thus, the second term in \eqref{eq:continuous_problem_1} promotes band-sparsity.
However, an attempt to solve \eqref{eq:continuous_problem_1} is confronted by two difficulties: 
(i) the regularization is not continuous, and
(ii) the optimization space is infinite-dimensional.

\noindent 
The first difficulty can be mitigated by relaxing \eqref{eq:continuous_problem_1} to 
\begin{align}
    \mathop{\textnormal{minimize}}_{\substack{\tilde{h}=\sum_{\bmn\in\setN}\tilde{h}_\bmn:\\ \textit{supp}(\underline{\tilde{h}}_\bmn)\in\setB_\bmn}}
    \sum_{\bmm\in\setM} \big| y(\bmz_\bmm) - \tilde{h}(\bmz_\bmm) \big|^2 + \sum_{\bmn\in\setN} r_{\bmn}(\|\tilde{h}_\bmn\|),
    \label{eq:continuous_problem_relaxed}
\end{align}
where the $r_{\bmn}$ are a continuous approximation to $\lambda_{\bmn}\mathbf{1}_{\opR_{>0}}$.
The second difficulty can be overcome by reducing \eqref{eq:continuous_problem_relaxed} to a finite-dimensional 
optimization problem using RKHS theory \cite{berlinet2011reproducing}. 

\subsection{RKHS Preliminaries}

A kernel is a function $K:\setZ\times\setZ\to\opC$ that is Hermitian, i.e., $K(\bmz,\bmz')=K(\bmz',\bmz)^\ast$, 
and positive semi-definite, i.e., 
\begin{align}
    \sum_{i=1}^{I}\sum_{i'=1}^{I} c_i c_{i'}^\ast K(\bmz_i,\bmz_{i'}) \geq 0
\end{align}
for all $I\in\opN$, $c_1,\dots, c_I\in\opC$, and $\bmz_1,\dots,\bmz_I\in\setZ$. Every such kernel 
corresponds to an RKHS, which is a space of functions defined as 
\begin{align}
    \setR(\setZ) \triangleq 
    \left\{\sum_{i=1}^I c_i\, K(\bmz_i,\cdot) \,\Big|\, I\in\opN, c_i\in\opC, \bmz_i\in\setZ  \right\}, 
\end{align}
that is endowed with the inner product
\begin{align}
    \!\!\bigg\langle\! \sum_{i=1}^I c_i\, K(\bmz_i,\cdot), \sum_{i=1}^I c'_i\, K(\bmz'_i,\cdot) \!\bigg\rangle
    \triangleq \sum_{i=1}^I c_i (c'_i)^\ast K(\bmz_i,\bmz'_i)\!
\end{align}
and the corresponding induced norm $\|\cdot\|\triangleq \sqrt{\langle\cdot,\cdot\rangle}$.
RKHSs owe their name to the so-called \emph{reproducing} property, according to which
evaluating $h\in \setR(\setZ)$ in $\bmz\in\setZ$ is equivalent to taking the inner product between $h$ and $K(\bmz,\cdot)$,
\begin{align}
    h(\bmz) = \big\langle h, K(\bmz,\cdot) \big\rangle.
\end{align}
The space of functions $h_{n_1}\!(\cdot)$ that satisfy \mbox{$\textit{supp}(\opF^{-1}\!\{h_n\})\!\subseteq\! \setT_{n_1}$}
(where $\setT_{n_1}$ is defined in \eqref{eq:T_interval}) is an RKHS with kernel 
\begin{align}
    &K_{n_1}^{\setT}(f,f')    \label{eq:define_K1} \\
    &= \frac{T}{N_1} \exp\!\bigg(\!2\pi i(f\!-\!f') \frac{({n_1}-\frac{N_1+1}{2})T}{N_1}\bigg) \sinc\!\bigg(\!\frac{(f\!-\!f')T}{N_1}\bigg). \nonumber
\end{align}
Moreover, if $K_1:\setZ_1\times\setZ_1\to\opC, (z_1,z_1')\mapsto K_1(z_1,z_1')$ 
and $K_2:\setZ_2\times\setZ_2\to\opC, (z_2,z_2')\mapsto K_2(z_2,z_2')$ are kernels, 
then $K:(\setZ_1\times\setZ_2)\times(\setZ_1\times\setZ_2)\to\opC, ((z_1,z_2),(z_1',z_2')) \mapsto K_1(z_1,z_1')K_2(z_2,z_2')$
is a kernel. Hence, the space of functions $h_{\bmn}(\bmz)\triangleq h_{\bmn}(f, x^{(\text{r})}, x^{(\text{c})})$ from \eqref{eq:h_atoms}
is an RKHS with kernel 
\begin{align}
    K_{\bmn}(\bmz, \bmz') &\triangleq K_{\bmn}\big((f, x^{(\text{r})}, x^{(\text{c})}), (f', x'^{(\text{r})}, x'^{(\text{c})})\big) \\
    &= K_{n_1}^{\setT}(f,f') K_{n_2}^{\setH}(x^{(\text{r})},x'^{(\text{r})}) K_{n_3}^{\setV}(x^{(\text{c})}, x'^{(\text{c})}),
\end{align}
where $K_{n_1}(f,f')$ is defined in \eqref{eq:define_K1}, and where
\begin{align}
    K_{n_2}^{\setH}(x^{(\text{r})}\!,x'^{(\text{r})})
    &\!=\! \frac{\Phi^{(\text{h})}}{N_2} \exp\!\bigg(\!2\pi i(x^{(\text{r})}\!\!-\!x'^{(\text{r})}) \frac{({n_2}-\frac{N_2+1}{2})\Phi^{(\text{h})}}{N_2}\!\bigg) 
    \nonumber \\
    &\hphantom{=}~\cdot \sinc\!\bigg(\!\frac{(x^{(\text{r})}\!-\!x'^{(\text{r})})\Phi^{(\text{h})}}{N_2}\bigg),
    \label{eq:define_K2} \\
    K_{n_3}^{\setV}(x^{(\text{c})}\!,x'^{(\text{c})}) 
    &\!=\! \frac{\Phi^{(\text{v})}}{N_3} \exp\!\bigg(\!2\pi i(x^{(\text{c})}\!\!-\!x'^{(\text{c})}) \frac{({n_3}-\frac{N_3+1}{2})\Phi^{(\text{v})}}{N_3}\!\bigg) 
    \nonumber \\
    &\hphantom{=}~\cdot \sinc\!\bigg(\!\frac{(x^{(\text{c})}\!-\!x'^{(\text{c})})\Phi^{(\text{v})}}{N_3}\bigg).
    \label{eq:define_K3}    
\end{align}

\vspace{-6.15mm}

Finally, if $\setR_1(\setZ)$ and $\setR_2(\setZ)$ are RKHSs with reproducing kernels $K_1$ and $K_2$ that satisfy 
$\setR_1(\setZ)\cap \setR_2(\setZ)=\{\boldsymbol{0}\}$, then 
$\setR(\setZ)=\{h=h_1+h_2 \,|\, h_1 \in \setR_1(\setZ), h_2 \in \setR_2(\setZ)\}$ is an RKHS
with reproducing kernel $K=K_1+K_2$, and its induced norm satisfies $\|h\|^2 = \|h_1\|^2 + \|h_2\|^2$ 
\cite[Sec. 1.4.1]{berlinet2011reproducing}. 

\subsection{Representer Theorem}
\label{sec:representer}

\begin{thm}[Representer Theorem] \label{thm:representer}
Consider a set $\setZ$ with RKHSs $\setR_{\bmn}(\setZ), \bmn\in\setN$
and corresponding reproducing kernels $K_{\bmn}$.
Suppose that we want to find a function $h=\sum_{\bmn\in\setN}h_{\bmn}$, with $h_{\bmn}\in\setR_{\bmn}$, that 
fits a collection of samples $(\bmz_\bmm, y_\bmm)\in\setZ\times\opC, \bmm\in\setM$,
by solving the regularized optimization problem 
\begin{align}
    \mathop{\textnormal{minimize}}_{\substack{\tilde{h}=\sum_{\bmn\in\setN}\tilde{h}_\bmn:\\\tilde{h}_\bmn\in\setR_\bmn(\setZ)}}
    \sum_{m=1}^M\big|y_\bmm - \tilde{h}(\bmz_\bmm)\big|^2 + \sum_{\bmn\in\setN}r_\bmn\big(\big\|\tilde{h}_\bmn\big\|\big), \label{eq:opt_prob}
\end{align}
where the $r_\bmn, \bmn\!\in\!\setN$ are increasing regularization functions. 
Then the solution $\hat{h}$ admits a representation of the form 
\begin{align}
    \hat{h}(\cdot) = \sum_{\bmn\in\setN}\sum_{\bmm\in\setM} c_{\bmn,\bmm}\, K_\bmn(\bmz_\bmm, \cdot), \label{eq:rep_sol}
\end{align}
where the coefficients $c_{\bmn,\bmm}$ are a solution for the problem
\begin{align}
    \!\!\!\mathop{\textnormal{minimize}}_{\substack{\tilde{c}_{\bmn,\bmm}:\\ \bmn\in\setN, \bmm\in\setM}}~~
    &\sum_{\bmm\in\setM}\big|y_\bmm - \sum_{\bmn\in\setN}\sum_{\bmm'\in\setM}\!\! \tilde{c}_{\bmn,\bmm'} K_\bmn(\bmz_{\bmm'},\bmz_\bmm) \big|^2 \nonumber\\
    + \sum_{\bmn\in\setN}& r_\bmn\Big(\!\big(\! \sum_{\bmm\in\setM}\sum_{\bmm'\in\setM}\!\! \tilde{c}_{\bmn,\bmm} \tilde{c}_{\bmn,\bmm'}^\ast K_\bmn(\bmz_{\bmm'},\bmz_\bmm) \big)^{\!\frac12}\Big).\! \label{eq:rep_coeff}
\end{align}
\end{thm}

The proof is in \fref{app:thm_proof}.
With \fref{thm:representer}, we can reduce the infinite-dimensional optimization problem in \eqref{eq:continuous_problem_relaxed}
to a finite-dimensional one. For this, we define the matrices 
\begin{align}
    \bK_{n_1}^{\setT}&\!(\Fpil) = \begin{bmatrix}
        K_{n_1}^{\setT}(f_1, f_1) & \!\!\!\cdots\!\!\! & K_{n_1}^{\setT}(f_{\Npil}, f_1)  \\
        \vdots & \!\!\!\ddots\!\!\! & \vdots \\
        K_{n_1}^{\setT}(f_1, f_{\Npil}) & \!\!\!\cdots\!\!\! & K_{n_1}^{\setT}(f_{\Npil}, f_{\Npil})
    \end{bmatrix} \!\! \in\! \opC^{\Npil\times\Npil} \label{eq:define_K1_matrix} \\
    \bK_{n_2}^{\setH} &= \begin{bmatrix}
        K_{n_2}^{\setH}(x^{(\text{r})}_1, x^{(\text{r})}_1) & \!\!\!\cdots\!\!\! & K_{n_2}^{\setH}(x^{(\text{r})}_{\Nrow}, x^{(\text{r})}_1)  \\
        \vdots & \!\!\!\ddots\!\!\! & \vdots \\
        K_{n_2}^{\setH}(x^{(\text{r})}_1, x^{(\text{r})}_{\Nrow}) & \!\!\!\cdots\!\!\! & K_{n_2}^{\setH}(x^{(\text{r})}_{\Nrow}, x^{(\text{r})}_{\Nrow})
    \end{bmatrix} \!\! \in \opC^{\Nrow\times\Nrow} \label{eq:define_K2_matrix} \\
    \bK_{n_3}^{\setV} &= \begin{bmatrix}
        K_{n_3}^{\setV}(x^{(\text{c})}_1, x^{(\text{c})}_1) & \!\!\!\cdots\!\!\! & K_{n_3}^{\setV}(x^{(\text{c})}_{\Ncol}, x^{(\text{c})}_1) \\
        \vdots & \!\!\!\ddots\!\!\! & \vdots \\
        K_{n_3}^{\setV}(x^{(\text{c})}_1, x^{(\text{c})}_{\Ncol}) & \!\!\!\cdots\!\!\! & K_{n_3}^{\setV}(x^{(\text{c})}_{\Ncol}, x^{(\text{c})}_{\Ncol})
    \end{bmatrix} \!\! \in \opC^{\Ncol\times\Ncol}
\end{align}
and
\begin{align}
    \bK_\bmn &= \bK_{n_1}^{\setT}\!(\Fpil) \kron \bK_{n_2}^{\setH} \kron \bK_{n_3}^{\setV} \in \opC^{(\Npil\Nrow\Ncol) \times (\Npil\Nrow\Ncol)},
    \label{eq:define_K_mat_n}
\end{align}
as well as the vectors 
\begin{align}
    \bmy &= [y(\bmz_\bmm)]_{\bmm\in\setM} \in \opC^{\Npil\Nrow\Ncol} \label{eq:define_y_vec}\\
    \tilde{\bmc}_{\bmn} &= [\tilde{c}_{\bmn,\bmm}]_{\bmm\in\setM} \in \opC^{\Npil\Nrow\Ncol} \label{eq:define_c_vec_m},
\end{align}
where the indices $\bmm$ are ordered such that they are consistent with the order of the entries of $\bK_\bmn$.

According to \fref{thm:representer}, we can obtain the solution to \eqref{eq:continuous_problem_relaxed} by solving
\begin{align}
    \{\hat\bmc_\bmn\}_{\bmn\in\setN} = 
    \mathop{\textnormal{argmin}}_{\tilde{\bmc}_{\bmn}:\,\bmn\in\setN}
    &\big\|\bmy -\!\! \sum_{\bmn\in\setN}\! \bK_\bmn \tilde{\bmc}_\bmn \big\|^2_2
   \!+\!\! \sum_{\bmn\in\setN}\! r_\bmn\big((\herm{\tilde{\bmc}_\bmn}\bK_\bmn \tilde{\bmc}_\bmn)^{\!\frac12}\big). \label{eq:rep_coeff_vec}
\end{align}
Note that the obtained coefficients $\{\hat\bmc_\bmn\}_{\bmn\in\setN}$ determine the the estimated function $\hat{h}$
according to \eqref{eq:rep_sol}, and that they do so for all $\bmz=(f, x^{(\text{r})}, x^{(\text{c})})\in\opR^3$ 
(not only for the measurument points $\{\bmz_\bmm\}_{\bmm\in\setM}=\Fpil\times\setX$). According to \fref{sec:problem}, 
we need to interpolate from the frequencies in $\Fpil$ to those in $\setF\supset\Fpil$. To this end, we also define the matrices
\begin{align}
    \bK_{n_1}^{\setT}\!(\setF;\Fpil) = \begin{bmatrix}
        K_{n_1}^{\setT}(f_1, f) & \!\!\!\!\!\cdots\!\!\!\!\! & K_{n_1}^{\setT}(f_{\Npil}, f)
    \end{bmatrix}_{f\in\setF} \!\in\! \opC^{|\setF|\times\Npil} \label{eq:define_K_freq_rec}
\end{align}
and
\begin{align}
    \bK_\bmn(\setF;\Fpil) &= \bK_{n_1}^{\setT}\!(\setF;\Fpil) \kron \bK_{n_2}^{\setH} \kron \bK_{n_3}^{\setV} \nonumber\\
    &\qquad \in \opC^{(|\setF|\Nrow\Ncol) \times (\Npil\Nrow\Ncol)}\!. \label{eq:define_Kmat_freq_n}
\end{align}
Given the coefficients $\{\hat\bmc_\bmn\}_{\bmn\in\setN}$ from \eqref{eq:rep_coeff_vec}, 
the channel estimate according to \eqref{eq:continuous_problem_relaxed} can now be evaluated in all points of interest by computing
\begin{align}
    \!\![\hat{h}(f,x^{(\text{r})},x^{(\text{c})})]_{(f,x^{(\text{r})},x^{(\text{c})})\in\setF\times\setX} 
    = \sum_{\bmn\in\setN} \bK_\bmn(\setF;\Fpil)\, \hat\bmc_\bmn.\!\!
\end{align}
To summarize:

\begin{mdframed}[style=csstyle,frametitle={RKHS-Regularized Channel Estimation}]
\vspace{-1mm}
The optimization problem in \eqref{eq:continuous_problem_relaxed} can be solved, and the obtained channel estimate 
$\hat{h}(f,x^{(\text{r})},x^{(\text{c})})$ can be evaluated at all frequencies $f\in\setF$ and antenna positions $(x^{(\text{r})},x^{(\text{c})})\in\setX$,
by solving 
\begin{align}
    \!\{\hat\bmc_\bmn\} \!=\! 
    \mathop{\textnormal{argmin}}_{\tilde{\bmc}_{\bmn}:\,\bmn\in\setN}
    &\big\|\bmy -\!\!\! \sum_{\bmn\in\setN} \bK_\bmn \tilde{\bmc}_\bmn \big\|^2_2 
    \!+\! \sum_{\bmn\in\setN}\! r_\bmn\big((\herm{\tilde{\bmc}_\bmn}\bK_\bmn \tilde{\bmc}_\bmn)^{\!\frac12}\big), \label{eq:rkhs_opt_summary}
\end{align}
where $\bK_\bmn$, $\bmy$, and $\tilde{\bmc}_\bmn$ are defined in \eqref{eq:define_K_mat_n}, \eqref{eq:define_y_vec}, and \eqref{eq:define_c_vec_m}, 
respectively, and by then computing
\begin{align}
    \![\hat{h}(f,x^{(\text{r})},x^{(\text{c})})]_{(f,x^{(\text{r})},x^{(\text{c})})\in\setF\times\setX} = \!\sum_{\bmn\in\setN} \bK_\bmn(\setF;\Fpil)\, \hat\bmc_\bmn, \label{eq:rkhs_opt_reconstruct}
\end{align}
where $\bK_\bmn(\setF)$ is defined in \eqref{eq:define_Kmat_freq_n}.
\end{mdframed}

\begin{remark}[Tikhonov Regularization]
\label{rem:tikhonov}
If the regularization functions $r_\bmn$ are chosen as
\begin{align}
    r_\bmn(x) = \gamma x^2, \quad \bmn\in\setN, 
\end{align}
then the regularization in \eqref{eq:continuous_problem_relaxed} is Tikhonov regularization \cite{tikhonov1977solutions}, 
$\gamma\sum_{\bmn\in\setN}\|\tilde{h}_\bmn\|^2 = \gamma\|\tilde{h}\|^2$, 
since $\setR_\bmn(\setZ)\cap\setR_{\bmn'}(\setZ)=\{\boldsymbol{0}\}$ for $\bmn\neq\bmn'$. 
\end{remark}

\begin{remark}[LASSO]
\label{rem:lasso}
If the regularization functions $r_\bmn$ are chosen as
\begin{align}
    r_\bmn(x) = \lambda x, \quad \bmn\in\setN, 
\end{align}
then \eqref{eq:continuous_problem_relaxed} is a LASSO problem \cite{tibshirani96} and \eqref{eq:rkhs_opt_summary} is a group LASSO problem \cite{yuan2006model}. 
\end{remark}

\begin{remark}[Elastic Net Regularization]
\label{rem:elastic}
The LASSO can also be combined with Tikhonov regularization
into elastic net regularization \cite{zou2005regularization} using regularization functions
\begin{align}
     r_\bmn(x) = \gamma x^2 + \lambda x, \quad \bmn\in\setN.
\end{align}
\end{remark}

\subsection{Dimensionality Reduction}
\label{sec:dimension}
To reduce the dimension of the latent variables $\tilde{\bmc}_\bmn$ (of which there are $|\setN|=N_1 N_2 N_3$ many) 
from $|\setM|=\Npil\Nrow\Ncol$, we use low-rank approximations of the kernel matrices. 
Recall that the $\bK_\bmn$ from \eqref{eq:define_K_mat_n} (which are used for solving \eqref{eq:rkhs_opt_summary}) 
are given as the Kronecker product of the three Hermitian positive semi-definite matrices $\bK_{n_1}^{\setT}\!(\Fpil)$, $\bK_{n_2}^{\setH}$, 
and $\bK_{n_3}^{\setV}$. 
Meanwhile, the $\bK_\bmn(\setF;\Fpil)$ from \eqref{eq:define_Kmat_freq_n} (which are used for inferring the interpolated
channel according to \eqref{eq:rkhs_opt_reconstruct}) are given as the Kronecker product of $\bK_{n_1}^{\setT}\!(\setF;\Fpil)$, 
which contains  $\bK_{n_1}^{\setT}\!(\Fpil)$ as a row subset, with the same $\bK_{n_2}^{\setH}$ and~$\bK_{n_3}^{\setV}$. 

For our low-rank approximations, we start by defining the \mbox{$|\setF|\times|\setF|$}-sized matrices 
\begin{align}
    \bK_{n_1}^{\setT}\!(\setF) = \begin{bmatrix}
        K_{n_1}^{\setT}(f, f')
    \end{bmatrix}_{f,f'\in\setF} \!\in\! \opC^{|\setF|\times|\setF|}.
\end{align}
Note that the $\bK_{n_1}^{\setT}\!(\Fpil)$ are (non-contiguous) submatrices of the $\bK_{n_1}^{\setT}\!(\setF)$ since $\Fpil\subset\setF$. 
We now perform the following eigenvalue decompositions (EVDs): 
\begin{align}
    \bK_{n_1}^{\setT}\!(\setF) &= \bV_{n_1}^{\setT}\!(\setF) \bLambda_{n_1}^{\setT}\!(\setF) \herm{(\bV_{n_1}^{\setT}\!(\setF))} \\
    \bK_{n_2}^{\setH} &= \bV_{n_2}^{\setH} \bLambda_{n_2}^{\setH} \herm{(\bV_{n_2}^{\setH})} \\
    \bK_{n_3}^{\setV} &= \bV_{n_3}^{\setV} \bLambda_{n_3}^{\setV} \herm{(\bV_{n_3}^{\setV})}.
\end{align}
We define $\bK_{n_1}^{\setT}\!(\Fpil;\setF)\in\opC^{\Npil\times|\setF|}$ and $\bV_{n_1}^{\setT}\!(\Fpil)\in\opC^{\Npil\times|\setF|}$
as the row-submatrices of $\bK_{n_1}^{\setT}\!(\setF)$ and $\bV_{n_1}^{\setT}\!(\setF)$, respectively, 
that consist of the rows which correspond to $\Fpil\subset\setF$. Hence, 
\begin{align}
    \bK_{n_1}^{\setT}\!(\Fpil) &= \bV_{n_1}^{\setT}\!(\Fpil) \bLambda_{n_1}^{\setT}\!(\setF) \herm{(\bV_{n_1}^{\setT}\!(\Fpil))} \\
    \bK_{n_1}^{\setT}\!(\setF;\Fpil) &= \bV_{n_1}^{\setT}\!(\setF) \bLambda_{n_1}^{\setF}\!(\setF) \herm{(\bV_{n_1}^{\setT}\!(\Fpil))}.
\end{align}
We therefore have the following decompositions of $\bK_\bmn$ and $\bK_\bmn(\setF;\Fpil)$ \cite[Thm.~4.2.12]{horn2011topics}:
\begin{align}
    \bK_\bmn
    &= \big(\bV_{n_1}^{\setT}\!(\Fpil) \kron \bV_{n_2}^{\setH} \kron \bV_{n_3}^{\setV}\big)
    \big(\bLambda_{n_1}^{\setT}\!(\setF) \kron \bLambda_{n_2}^{\setH} \kron \bLambda_{n_3}^{\setV} \big) \qquad \nonumber \\
    &\quad~ \cdot \herm{\big(\bV_{n_1}^{\setT}\!(\Fpil) \kron \bV_{n_2}^{\setH} \kron \bV_{n_3}^{\setV} \big)}
\end{align}
and
\begin{align}
    \bK_\bmn(\setF;\Fpil)
    &= \big(\bV_{n_1}^{\setT}\!(\setF) \kron\! \bV_{n_2}^{\setH} \!\kron\! \bV_{n_3}^{\setV}\big)
    \big(\bLambda_{n_1}^{\setT}\!(\setF) \kron\! \bLambda_{n_2}^{\setH} \!\kron\! \bLambda_{n_3}^{\setV} \big) \nonumber \\
    &\quad~ \cdot \herm{\big(\bV_{n_1}^{\setT}\!(\Fpil) \kron \bV_{n_2}^{\setH} \kron \bV_{n_3}^{\setV} \big)}.    
\end{align}

We form low-rank approximations of $\bK_\bmn$ and $\bK_\bmn(\setF;\Fpil)$ by retaining only the $R$ 
dominant ranks of $\bK_{n_1}^{\setT}\!(\Fpil)$, $\bK_{n_1}^{\setT}\!(\setF;\Fpil)$, $\bK_{n_2}^{\setH}$, and $\bK_{n_3}^{\setV}$:\footnote{
In principle, one could retain a different number of ranks for these different matrices. 
In a perfunctory attempt to salvage any semblance of simplicity, we use the same rank number for all matrices.
}
\begin{align}
    \!\!\!\bK_{n_1}^{\setT}\!(\Fpil) &\approx \!
    \sum_{r=1}^R [\lambda_{n_1}^{\setT}(\setF)](r)\, [\bmv_{n_1}^{\setT}(\Fpil)](r)\, \herm{\big([\bmv_{n_1}^{\setT}(\Fpil)](r)\big)}\!\! \\
    &= \bS_{n_1}^{\setT}\!(\Fpil) \herm{\big(\bS_{n_1}^{\setT}\!(\Fpil)\big)}, \label{eq:lr_KT_Fpil} \\
    \!\!\!\bK_{n_1}^{\setT}\!(\setF; \Fpil) &\approx \!
    \sum_{r=1}^R [\lambda_{n_1}^{\setT}(\setF)](r)\, [\bmv_{n_1}^{\setT}(\setF)](r)\, \herm{\big([\bmv_{n_1}^{\setT}(\Fpil)](r)\big)}\!\! \\
    &= \bS_{n_1}^{\setT}\!(\setF) \herm{\big(\bS_{n_1}^{\setT}\!(\Fpil)\big)}, \label{eq:lr_KT_setF}
\end{align}
where 
\begin{align}
    \!\!\!\bS_{n_1}^{\setT}\!(\Fpil)
    \triangleq& \Big[\sqrt{[\lambda_{n_1}^{\setT}(\setF)](1)}\,[\bmv_{n_1}^{\setT}(\Fpil)](1), \,\dots, \nonumber\\
     &\quad \sqrt{[\lambda_{n_1}^{\setT}(\setF)](R)}\,[\bmv_{n_1}^{\setT}(\Fpil)](R)\Big]\!\in \opC^{\Npil\times R},\!\! \\
    \!\!\!\bS_{n_1}^{\setT}\!(\setF)
    \triangleq& \Big[\sqrt{[\lambda_{n_1}^{\setT}(\setF)](1)}\,[\bmv_{n_1}^{\setT}(\setF)](1), \,\dots, \nonumber\\
     &\quad \sqrt{[\lambda_{n_1}^{\setT}(\setF)](R)}\,[\bmv_{n_1}^{\setT}(\setF)](R)\Big]\!\in \opC^{|\setF|\times R},\!\!
\end{align}
and analogously
\begin{align}
    \bK_{n_2}^{\setH} &\approx 
    \sum_{r=1}^R \lambda_{n_2}^{\setH}(r)\, \bmv_{n_2}^{\setH}(r)\, \herm{\big(\bmv_{n_2}^{\setH}(r)\big)}\!\!
    = \bS_{n_2}^{\setH} \herm{\big(\bS_{n_2}^{\setH}\big)},  \label{eq:lr_KH} \\
    \bK_{n_3}^{\setV} &\approx 
    \sum_{r=1}^R \lambda_{n_3}^{\setV}(r)\, \bmv_{n_3}^{\setV}(r)\, \herm{\big(\bmv_{n_3}^{\setV}(r)\big)}\!\!
    = \bS_{n_3}^{\setV} \herm{\big(\bS_{n_3}^{\setV}\big)}.  \label{eq:lr_KV} 
\end{align}
The low-rank approximations \eqref{eq:lr_KT_Fpil}, \eqref{eq:lr_KT_setF}, \eqref{eq:lr_KH}, and \eqref{eq:lr_KV}
give us the following low-rank approximations for $\bK_\bmn$ and $\bK_\bmn(\setF;\Fpil)$: 
\vspace{-1mm}
\begin{align}
    \!\bK_\bmn &\approx 
    \underbrace{\big(\bS_{n_1}^{\setT}\!(\Fpil) \!\otimes\! \bS_{n_2}^{\setH} \!\otimes\! \bS_{n_3}^{\setV}\big)}_{\triangleq \bS_\bmn  \in \opC^{(\Npil\Nrow\Ncol)\times R^3}\hspace{-2.7cm}}
    \herm{\big(\bS_{n_1}^{\setT}\!(\Fpil) \!\otimes\! \bS_{n_2}^{\setH} \!\otimes\! \bS_{n_3}^{\setV}\big)}
    \label{eq:define_bS}\! \\
    &= \bS_\bmn \herm{\bS_\bmn}
\end{align}
and
\begin{align}
    \bK_\bmn(\setF;\Fpil) &\approx 
    \underbrace{\big(\bS_{n_1}^{\setT}\!(\setF) \!\otimes\! \bS_{n_2}^{\setH} \!\otimes\! \bS_{n_3}^{\setV}\big)}_{\triangleq \bS_\bmn(\setF)  \in \opC^{(|\setF|\Nrow\Ncol)\times R^3}\hspace{-3.15cm}}
    \herm{\bS_\bmn} \label{eq:define_bS_F} \\
    &= \bS_\bmn(\setF)\, \herm{\bS_\bmn}.
\end{align}
We can now define $\tilde{\bma}_{\bmn}\triangleq \herm{\bS_\bmn}\tilde{\bmc}_{\bmn}\in\opC^{R^3}$
and replace \eqref{eq:rkhs_opt_summary} and \eqref{eq:rkhs_opt_reconstruct} 
with the following optimization problem, 
for which the dimension of the optimization variables $\tilde{\bma}_\bmn$ (instead of $\tilde{\bmc}_\bmn$) 
is reduced to $R^3$ (compared to $\Npil\Nrow\Ncol$):

\begin{mdframed}[style=csstyle,frametitle={Low-Rank RKHS-Regularized Channel Estimation}]
\vspace{-1mm}
A low-rank surrogate to the optimization problem in \eqref{eq:continuous_problem_relaxed} 
can be solved, and the obtained channel estimate $\hat{h}(f,x^{(\text{r})},x^{(\text{c})})$ can be evaluated 
at all frequencies $f\in\setF$ and antenna positions $(x^{(\text{r})},x^{(\text{c})})\in\setX$, by solving 
\begin{align}
    \!&\{\hat\bma_\bmn\}_{\bmn\in\setN} \!=\! 
    \mathop{\textnormal{argmin}}_{\tilde{\bma}_{\bmn}:\,\bmn\in\setN}
    \big\|\bmy -\!\!\! \sum_{\bmn\in\setN} \bS_\bmn \tilde{\bma}_\bmn \big\|^2_2  
    \!+\! \sum_{\bmn\in\setN}\! r_\bmn(\|\tilde{\bma}_{\bmn}\|_2),\!  \label{eq:lr_optimization_problem} \\
    \!&[\hat{h}(f,x^{(\text{r})},x^{(\text{c})})]_{(f,x^{(\text{r})},x^{(\text{c})})\in\setF\times\setX} = \!\sum_{\bmn\in\setN} \bS_\bmn(\setF)\, \hat\bma_\bmn,  \label{eq:lr_reconstruction}
\end{align}
where $\bS_\bmn$ and $\bS_\bmn(\setF)$ are defined in \eqref{eq:define_bS} and \eqref{eq:define_bS_F}. 
\end{mdframed}
\noindent
Note that the operation $\sum_{\bmn\in\setN} \bS_\bmn \tilde{\bma}_\bmn$ in \eqref{eq:lr_optimization_problem}
can be interpreted as matrix-vector product between the \mbox{$(\Npil\Nrow\Ncol)\times (R^3|\setN|)$} matrix $\bsfS$,\footnote{Remember
that $|\setN|=N_1 N_2 N_3$, see \fref{sec:model}.
}
whose column blocks are the $\bS_\bmn$ for all $\bmn\in\setN$, and the $(R^3|\setN|)$-dimensional vector $\tilde{\bsfa}$
obtained by stacking the $\tilde{\bma}_\bmn$ for all $\bmn\in\setN$.\footnote{
Analogously, the operation $\sum_{\bmn\in\setN} \bS_\bmn(\setF) \tilde{\bma}_\bmn$ in \eqref{eq:lr_reconstruction}
can also be interpreted as one big matrix-vector product between a matrix $\bsfS(\setF)$, whose column blocks are the $\bS_\bmn(\setF)$
for all $\bmn\in\setN$, and the vector $\tilde\bsfa$.
}
The optimization algorithm for solving \eqref{eq:lr_optimization_problem} (see \fref{sec:algorithm}) will have to repeatedly 
perform matrix-vector multiplications involving $\bsfS$ as well as its Hermitian transpose $\herm{\bsfS}$.
Na\"ively, each of these operations involves $R^3\Npil\Nrow\Ncol|\setN|$ scalar multiplications. 
However, we now show how to exploit the modulation structure of the kernels and perform these operations with just 
$O(R^3 |\setN| \log|\setN|)$ scalar multiplications by using fast Fourier algorithms. 

\subsection{Modulation Structure}
\label{sec:modulation}

Note that, for the different box indices $\bmn$, the kernel matrices $\bK_{n_1}^{\setT}(\setF)$, 
$\bK_{n_2}^{\setH}$, and $\bK_{n_3}^{\setV}$ change only by complex modulation. 
This enables us to express the eigenvectors of these matrices as modulations of common sets of eigenvectors. 
For the sake of brevity, we show only the case of the kernel matrix $\bK_{n_2}^{\setH}$. 
We start by defining the indexless (=no dependence on~$n_2$)~kernel 
\begin{align}
    K^{\setH}(x^{(\text{r})}\!,x'^{(\text{r})})
    &\!=\! \frac{\Phi^{(\text{h})}}{N_2} \sinc\!\bigg(\!\frac{(x^{(\text{r})}\!-\!x'^{(\text{r})})\Phi^{(\text{h})}}{N_2}\bigg),
\end{align}
which differs from \eqref{eq:define_K2} by its omission of the complex sinusoid. 
We then define the corresponding indexless kernel matrix $\bK^{\setH}\in\opC^{\Nrow\times\Nrow}$ analogous to \eqref{eq:define_K2_matrix}. 
Moreover, we define the complex wave vector 
\begin{align}
    \!\!\bmf_{n_2}^{\setH} =&
    \tp{\bigg[\!\exp\!\bigg(\!2\pi i x^{(\text{r})}_1 \frac{(n_2-\frac{N_2+1}{2})\Phi^{(\text{h})}}{N_2} \bigg), \dots, \nonumber\\
    &\qquad \exp\!\bigg(\!2\pi i x^{(\text{r})}_{\Nrow} \frac{(n_2-\frac{N_2+1}{2})\Phi^{(\text{h})}}{N_2} \bigg) \bigg]} \in \opC^{\Nrow}.
\end{align}
The following now holds (the proof is in \fref{app:lem_proof}):
\begin{lem}\label{lem:phase}
Let $\bmv$ be an eigenvector of $\bK^{\setH}$ with corresponding eigenvalue $\lambda$. 
Then $\bmv_{n_2}=\bmf_{n_2}^{\setH}\odot\bmv$ is an eigenvector of $\bK_{n_2}^{\setH}$ with identical corresponding eigenvalue 
$\lambda_{n_2}=\lambda$.
\end{lem}

Analogous results hold for suitably defined $\bK^{\setT}(\setF)$, $\bK^{\setT}(\Fpil)$, and $\bK^{\setV}$, 
and $\bmf_{n_1}^{\setT}(\setF)$, $\bmf_{n_1}^{\setT}(\Fpil)$, and $\bmf_{n_3}^{\setV}$.

\fref{lem:phase} enables us to compute the matrix-vector product with $\bsfS$ and with $\herm{\bsfS}$ 
more efficiently (as well as the analogous matrix-vector product in \eqref{eq:lr_reconstruction}).
To this end, recall the low-rank approximations of the kernel matrices in \fref{sec:dimension}. 
We now define analogous low-rank approximations of the indexless kernel matrices $\bK^{\setT}(\setF)$, $\bK^{\setT}(\Fpil)$, 
$\bK^{\setH}$, and $\bK^{\setV}$:
\begin{align}
    \bK^{\setT}(\setF) &\approx \bS^{\setT}(\setF) \herm{(\bS^{\setT}(\setF))}, \\
    \bK^{\setT}(\Fpil) &\approx \bS^{\setT}(\Fpil) \herm{(\bS^{\setT}(\Fpil))}, \\
    \bK^{\setH} &\approx \bS^{\setH} \herm{(\bS^{\setH})}, \\
    \bK^{\setV} &\approx \bS^{\setV} \herm{(\bS^{\setV})}.
\end{align}
From \fref{lem:phase}, it follows that 
\begin{align}
    \bS_\bmn =&~ \bS_{n_1}^{\setT}\!(\Fpil) \!\otimes\! \bS_{n_2}^{\setH} \!\otimes\! \bS_{n_3}^{\setV} \\
    =&~ \Big(\bS^{\setT}\!(\Fpil) \odot \big(\bmf_{n_1}^{\setT}\!(\Fpil) \tp{\mathds{1}_R}\big)\Big) \otimes \Big(\bS^{\setH} \odot \big( \bmf_{n_2}^{\setH} \tp{\mathds{1}_R}\big)\Big) \nonumber \\
    &~\otimes \Big(\bS^{\setV} \odot \big(\bmf_{n_3}^{\setV} \tp{\mathds{1}_R}\big)\Big) \\
    =&~ \Big(\bS^{\setT}\!(\Fpil) \otimes \bS^{\setH} \otimes \bS^{\setV} \Big) \nonumber \\
    &~\odot \Big( \big(\bmf_{n_1}^{\setT}\!(\Fpil) \tp{\mathds{1}_R} \big) \otimes \big(\bmf_{n_2}^{\setH} \tp{\mathds{1}_R}\big) \otimes \big(\bmf_{n_3}^{\setV} \tp{\mathds{1}_R}\big) \Big) \\
    =&~ \underbrace{\big(\bS^{\setT}\!(\Fpil) \otimes \bS^{\setH} \otimes \bS^{\setV} \big)}_{\triangleq \bS \in \opC^{(\Npil\Nrow\Ncol)\times R^3}\hspace{-2.55cm}} 
    \odot \underbrace{\big(\bmf_{n_1}^{\setT}\!(\Fpil) \otimes \bmf_{n_2}^{\setH} \otimes \bmf_{n_3}^{\setV}\big)}_{\triangleq \bmf_\bmn \in \opC^{\Npil\Nrow\Ncol}\hspace{-1.95cm}} \tp{\mathds{1}_{R^3}} \\
    =&~ \bS \odot \big(\bmf_\bmn\,\tp{\mathds{1}_{R^3}}\big).
\end{align}
We can therefore compute the matrix vector product with $\bsfS$, as in \eqref{eq:lr_optimization_problem}, as
\begin{align}
    \bsfS \tilde\bsfa &= \sum_{\bmn\in\setN} \bS_\bmn \tilde{\bma}_\bmn
    = \sum_{\bmn\in\setN} \Big( \bS \odot \big(\bmf_\bmn\,\tp{\mathds{1}_{R^3}}\big) \Big) \tilde{\bma}_\bmn \\
    &= \big(\bS \odot \bF\tilde{\bA}\big) \mathds{1}_{\Npil\Nrow\Ncol}, \label{eq:product_with_bsfS}
\end{align}
where we have defined the $(\Npil\Nrow\Ncol)\times |\setN|$ Fourier matrix $\bF$, whose columns are the $\bmf_\bmn$, $\bmn\in\setN$, 
as well as the $|\setN|\times R^3$ coefficient matrix $\tilde{\bA}$, whose rows are the $\tp{\tilde{\bma}_\bmn}$, $\bmn\in\setN$.

Analogously, \eqref{eq:lr_reconstruction} can be computed as 
\begin{align}
    \sum_{\bmn\in\setN} \bS_\bmn(\setF)\, \tilde{\bma}_\bmn 
    = \big(\bS(\setF) \odot \bF\hat{\bA}\big) \mathds{1}_{\Npil\Nrow\Ncol},
\end{align}
where $\bS(\setF)\triangleq \bS^{\setT}\!(\setF) \otimes \bS^{\setH} \otimes \bS^{\setV}$, 
and where the rows of $\hat{\bA}$ are the solution coefficients $\hat{\bma}_\bmn$ from \eqref{eq:lr_optimization_problem}.

The product between $\herm{\bsfS}$ and an $(\Npil\Nrow\Ncol)$-dimensional vector $\bme$
(which will be needed in \fref{sec:algorithm} below) can be computed analogously as 
\begin{align}
    \herm{\bsfS} \bme &= \Big[\herm{\bS_\bmn} \Big]_{\bmn\in\setN} \bme \\
    &= \Big[\herm{(\bmf_\bmn\,\tp{\mathds{1}_{R^3}})} \odot \herm{\bS} \Big]_{\bmn\in\setN} \bme \\
    &=  \Big[\big(\mathds{1}_{R^3}\herm{\bmf_\bmn}\big) \odot \herm{\bS} \Big]_{\bmn\in\setN} \bme \\
    &= \text{vec}\Big( \tp{\big(\herm{\bF} \big(\bS^\ast \odot (\bme\tp{\mathds{1}_{R^3}}) \big) \big)}\, \Big). \label{eq:product_with_herm_bsfS}
\end{align}
\vspace{-6mm}

\subsection{Optimization Algorithm}
\label{sec:algorithm}
We now provide an algorithm to solve  \eqref{eq:lr_optimization_problem} for the case of elastic net regularization, 
i.e., for $r_\bmn(x)=\gamma x^2 + \lambda x$, $\bmn\in\setN$. The Tikhonov regularization case and the LASSO
case follow as special cases with $\lambda=0$ and $\gamma=0$, respectively. 
We start by writing down the objective of \eqref{eq:lr_optimization_problem} explicitly: 
\begin{align}
    \big\|\bmy - \bsfS \tilde\bsfa \big\|^2_2  
    + \gamma \|\tilde\bsfa\|^2_2 + \lambda \sum_{\bmn\in\setN} \|\tilde{\bma}_\bmn\|_2. \label{eq:elastic_net_opt}    
\end{align}
For optimization, we use the FBS framework \cite{goldstein16a}, also known as proximal gradient descent. 
FBS iteratively solves problems of the form 
\vspace{-2mm}
\begin{align}
    \mathop{\textnormal{minimize}}_{\tilde\bmx} f(\tilde\bmx) + g(\tilde\bmx), \label{eq:basic_fbs}
\end{align}
where $f$ is convex\footnote{FBS can also be employed when $f$ is non-convex, but in that case, there are typically no guarantees for convergence or optimality.} and differentiable, and $g$ is convex. 
FBS solves such problems by starting at some initial position $\tilde\bmx^{(0)}$ and then alternating the following two steps:
the \emph{descent} step
\begin{align}
    \breve\bmx^{(t)} = \tilde\bmx^{(t)} - \tau^{(t)} \, \nabla_{\tilde\bmx} f\big|_{\tilde\bmx=\tilde\bmx^{(t)}} \label{eq:fbs_descent}
\end{align}
and the \emph{proximal} step
\begin{align}
    \tilde\bmx^{(t+1)} = \argmin_{\tilde\bmx} \tau^{(t)} g(\tilde\bmx) + \frac12 \big\|\tilde\bmx - \breve\bmx^{(t)}\big\|^2_2. \label{eq:fbs_prox}
\end{align}
Given a suitable choice of stepsizes\footnote{A sufficient condition is for all stepsizes to be bounded away from zero 
while being smaller than the reciprocal of the Lipschitz constant of $\nabla f$.}
$\tau^{(0)}, \tau^{(1)},\dots$, FBS is guaranteed to converge to the optimal solution. 
In practice, it is common to terminate the algorithm after a fixed number of iterations $t_{\max}$ regardless of whether or not it has converged.

We can leverage FBS for our problem by identifying
$f$ with $\big\|\bmy - \bsfS\tilde\bsfa \big\|^2_2 + \gamma \|\tilde\bsfa\|^2_2$
and $g$ with $\lambda \sum_{\bmn\in\setN} \|\tilde{\bma}_\bmn\|_2$. 
In other words, $f$ is the quadratic part of the objective, and $g$ is the non-quadratic part.
In that case, the descent step is
\begin{align}
    \breve\bsfa^{(t)} &= \tilde\bsfa^{(t)} - \tau^{(t)} \Big(  \herm\bsfS(\bsfS\tilde\bsfa^{(t)} - \bmy) + 2\gamma\, \tilde\bsfa^{(t)} \Big),
    \label{eq:fbs_explicit_descent}
\end{align}
where the matrix-vector products with $\bsfS$ and $\herm\bsfS$ can be efficiently computed using \eqref{eq:product_with_bsfS} and \eqref{eq:product_with_herm_bsfS}, respectively.
The proximal step amounts to group-wise soft-thresholding:
\begin{align}
    \tilde\bma_\bmn^{(t+1)} = \frac{\breve\bma_\bmn^{(t)}}{\|\breve\bma_\bmn^{(t)}\|_2} \max\big\{\|\breve\bma_\bmn^{(t)}\|_2 - \tau^{(t)}\lambda, 0 \big\}, ~ \bmn\in\setN, \label{eq:fbs_explicit_prox}
\end{align}
where the $\breve\bma_\bmn^{(t)}$ are the groups of $\breve\bsfa^{(t)}$ (in the same sense as the $\tilde\bma_\bmn$ are the groups of $\tilde\bsfa$). 
A natural initialization is $\tilde\bsfa^{(0)}=\herm\bsfS\bmy$. 

\subsection{Debiasing}
\label{sec:debiasing}
Recall that the purpose of our regularization was to exploit the intrinsic band-sparsity of wireless channels in delay-beamspace. 
A natural way of doing this while still preserving the convexity of the objective is to use LASSO regularization (see Remark \ref{rem:lasso})
or elastic net regularization (see Remark \ref{rem:elastic}) with $\lambda\gg0$. 
However, while such regularization succeeds at delivering a band-sparse channel estimate, it has the undesirable side-effect of 
introducing a substantial bias (towards zero) even in the channel atoms $\bmn$ found to be significant. 

One way to alleviate this is to not estimate the channel directly using \eqref{eq:lr_reconstruction},
but to instead only extract the found support $\setS\subset\setN$ from \eqref{eq:lr_optimization_problem} as
\begin{align}
    \setS = \{\bmn\in\setN \,|\, \hat\bma_\bmn \neq \boldsymbol{0}\}, \label{eq:define_support_set}
\end{align}
where $\{ \hat\bma_\bmn\}_{\bmn\in\setN}$ is the solution to \eqref{eq:lr_optimization_problem}
and to then perform Tikhonov-regularized\footnote{\label{footnote:tikhonov}We retain Tikhonov regularization to prevent overfitting since
even a single atom $\tilde{h}_\bmn$ in \eqref{eq:opt_prob} could perfectly interpolate the measurements, which would 
result in plain-vanilla least-squares channel estimation. (The low-rank approximation from \fref{sec:dimension} amounts 
to a form of implicit regularization that already tends to prevent this---but we prefer the additional use of a
more principled regularization.)}
channel estimation \emph{in the subspace spanned by the atoms of $\setS$}, 
i.e., to then solve
\begin{align}
    \mathop{\textnormal{minimize}}_{\substack{\tilde{h}=\sum_{\bmn\in\textcolor{myred}{\setS}}\tilde{h}_\bmn:\\ \textit{supp}(\underline{\tilde{h}}_\bmn)\in\setB_\bmn}}
    \sum_{\bmm\in\setM} \big| y(\bmz_\bmm) - \tilde{h}(\bmz_\bmm) \big|^2 + \frac{1}{\textit{SNR}} \sum_{\bmn\in\textcolor{myred}{\setS}} \|\tilde{h}_\bmn\|^2,
\end{align}
instead of \eqref{eq:continuous_problem_relaxed}. The key difference to \eqref{eq:continuous_problem_relaxed} is highlighted in red, 
and the regularizer functions are $r_\bmn(x)=(1/\textit{SNR})x^2$, $\bmn\in\setN$, as in Remark \ref{rem:elastic}, were $\textit{SNR}$ represents 
the received signal-to-noise ratio (SNR). 

Since a subspace of an RKHS is itself an RKHS, \fref{thm:representer} can be applied by substituting $\setN$
with $\setS$. One can then use the same dimensionality-reduction techniques as in \fref{sec:dimension} to obtain the following debiasing problem:

\begin{mdframed}[style=csstyle,frametitle={Debiasing the RKHS-Regularized Channel Estimate}]
\vspace{-1mm}
A debiased sparse channel estimate can be obtained by solving 
\begin{align}
    \!\!&\{\hat\bma_\bmn\}_{\bmn\in\setS} \!=\! 
    \mathop{\textnormal{argmin}}_{\tilde{\bma}_{\bmn}:\,\bmn\in\setS}
    \big\|\bmy -\!\!\! \sum_{\bmn\in\setS} \bS_\bmn \tilde{\bma}_\bmn \big\|^2_2  
    \!+\! \frac{1}{\textit{SNR}} \sum_{\bmn\in\setS}\! \|\tilde{\bma}_{\bmn}\|^2_2,\!\!  \label{eq:debiasing_optimization_problem} \\
    \!&[\hat{h}(f,x^{(\text{r})},x^{(\text{c})})]_{(f,x^{(\text{r})},x^{(\text{c})})\in\setF\times\setX} = \!\sum_{\bmn\in\setS} \bS_\bmn(\setF)\, \hat\bma_\bmn,  \label{eq:debiasing_reconstruction}
\end{align}
where $\bS_\bmn$ and $\bS_\bmn(\setF)$ are defined in \eqref{eq:define_bS} and \eqref{eq:define_bS_F},~and where $\setS$ is defined in \eqref{eq:define_support_set}. A good value for the regularization constant $\alpha$ is the reciprocal of the 
receive~SNR.
\end{mdframed}

\noindent
The same modulation structure as in \fref{sec:modulation} can now be exploited, and an FBS algorithm analogous to the one 
in \fref{sec:algorithm} can be used by keeping \eqref{eq:fbs_explicit_descent} (with $1/\textit{SNR}$ taking the place of~$\gamma$), but by replacing \eqref{eq:fbs_explicit_prox} with\footnote{
This algorithm is obtained by choosing the function $g$ from \eqref{eq:basic_fbs} to be 
$g(\tilde\bsfa)=\sum_{\bmn\in\setN\setminus\setS}g'(\|\tilde\bma_\bmn\|^2_2)$, 
where $g'(x)=0$ if $x=0$ and $\infty$ otherwise,
and by keeping the same function $f$ as in \fref{sec:algorithm}.
}
\begin{align}
    \tilde\bma_\bmn^{(t+1)} = \begin{cases} 
    \breve\bma_\bmn^{(t)}   &\text{for}~ \bmn\in\setS \\
    \boldsymbol{0} &\text{for}~ \bmn\in\setN\setminus\setS.
    \end{cases}
\end{align}
We can now put together all the pieces and obtain the channel estimation 
algorithm summarized in \fref{alg:rkhs}.

\subsection{Computational Complexity}
The computational complexity of the descent steps in lines 5 and 11 of \fref{alg:rkhs} is 
dominated by the matrix-vector products $\bsfS\tilde\bsfa^{(t)}$ and $\herm\bsfS(\bsfS\tilde\bsfa^{(t)}\!-\bmy)$, 
which can be computed using fast Fourier algorithms with $O(R^3 |\setN| \log|\setN|)$ complex floating-point operations 
(see \fref{sec:modulation}). The same holds for the initialization in line 3 and the channel reconstruction in line 15. 
The proximal operator in line 6 requires the norm computation and rescaling of all atoms $\tilde\bma_{\bmn}^{(t)}$. 
Both of these operations require a total of $O(|\setN|R^3)$ complex floating-point operations (assuming that a square
root can be computed in $O(1)$ floating-point operations). 
The computational complexity of the proximal operator in line 12 is negligible. 
Hence, the total complexity of \fref{alg:rkhs} is $O((t_{\max}\!+t'_{\max})R^3 |\setN| \log|\setN|)$. 

\begin{algorithm}[tp]
  \caption{RKHS Channel Estimation}
  \label{alg:rkhs}
  \begin{algorithmic}[1]
    \Function{RKHS-CHEST}{$\bmy$}
    \vspace{1mm}
    \State \textcolor{myaltgray}{//\,sparse optimization (see \fref{sec:algorithm})}\vspace{0.2mm}
    \State $\tilde\bsfa^{(0)}=\herm\bsfS\bmy$
    \For{$t=0,1,\dots,t_{\max}-1$}
        \State $\breve\bsfa^{(t)} = \tilde\bsfa^{(t)} - \tau^{(t)} \big(  \herm\bsfS(\bsfS\tilde\bsfa^{(t)} - \bmy) + 2\gamma\, \tilde\bsfa^{(t)} \big)$
        \State $\tilde\bma_\bmn^{(t+1)}\!=\! (\breve\bma_\bmn^{(t)}\!/\|\breve\bma_\bmn^{(t)}\|_2) \max\big\{\!\|\breve\bma_\bmn^{(t)}\|_2 \!- \tau^{(t)}\lambda, 0 \big\}, \bmn\!\in\!\setN$
    \EndFor
    \vspace{3mm}
    \State \textcolor{myaltgray}{//\,debiasing (see \fref{sec:debiasing})}\vspace{0.2mm}
    \State $\setS = \{\bmn\in\setN \,|\, \tilde\bma_\bmn^{(t_{\max})} \neq \boldsymbol{0}\}$
    \For{$t=t_{\max},\dots,t_{\max}\!+t'_{\max}\!-\!1$}\vspace{1mm}
        \State $\breve\bsfa^{(t)} = \tilde\bsfa^{(t)} - \tau^{(t)} \big(  \herm\bsfS(\bsfS\tilde\bsfa^{(t)} - \bmy) + (2/\textit{SNR})\, \tilde\bsfa^{(t)} \big)$
        \State $    \tilde\bma_\bmn^{(t+1)} = \begin{cases} 
    \breve\bma_\bmn^{(t)}   &\text{for}~ \bmn\in\setS \\
    \boldsymbol{0} &\text{for}~ \bmn\in\setN\setminus\setS.
    \end{cases}$\vspace{-0.5mm}
    \EndFor
    \vspace{3mm}
    \State \textcolor{myaltgray}{//\,function evaluation}\vspace{0.2mm}
    \State \Return $\bsfS(\setF)\,\tilde\bsfa^{(t_{\max}\!+t'_{\max})}$
    \vspace{1mm}
    \EndFunction    
  \end{algorithmic}
\end{algorithm}

\section{Data-Driven Enhancements}
\label{sec:deep_unfolding}

In \fref{sec:chest}, we have proposed a method for channel estimation that exploits the band-sparsity
of wireless channel estimators in the delay-beamspace domain through regularization in an RKHS. 
The actual channel estimator boils down to the iterative optimization algorithm from \fref{alg:rkhs}.

However, that algorithm depends on a number of hyperparameters that need tuning, such as the 
regularization constants~$\lambda$ and/or $\gamma$ (depending on which type of regularization is used)
and the stepsizes $\tau^{(k)}$. 
Moreover, not all channel atoms might have the same predisposition to be zero. Consider that the different atoms correspond to different 
channel delays and different angles of arrival (both in azimuth and elevation). Early delays might be much 
more likely to carry energy than late ones; vice versa, elevation angles within the plane might be much more
likely to carry energy than those from above. Thus, it seems reasonable to account for these differences with
anisotropic regularization, where the regularization constants $\lambda_\bmn$ and/or $\gamma_\bmn$ depend on 
the atom index $\bmn$. This leads to an explosion in the number of tunable hyperparameters, however. 

To manage such a large number of hyperparameters, we therefore use deep unfolding \cite{balatsoukas2019deep}, 
where the iterations and hyperparameters of an algorithm are interpreted as the layers and trainable weights of a deep neural network, 
respectively. Data-driven methods based on digital twinning or self-supervised techniques can then be used for learning
these weights using stochastic gradient descent and back-propagation.\footnote{The development of a self-supervised method
is left for future work.}

An obstacle for using \fref{alg:rkhs} in this way is the following: Assuming that the debiasing
stage (lines 9 to 13) uses a sufficient number of suitable stepsizes to converge, then the output of \fref{alg:rkhs} depends on 
the trainable parameters of the sparse optimization stage (lines 3 to 7) only through the obtained support set $\setS$. 
But this set is not differentiable with respect to the trainable parameters, so that these parameters cannot be effectively 
learned through backpropagation. 

\begin{figure}[tp]
\centering
\includegraphics[width=0.9\columnwidth]{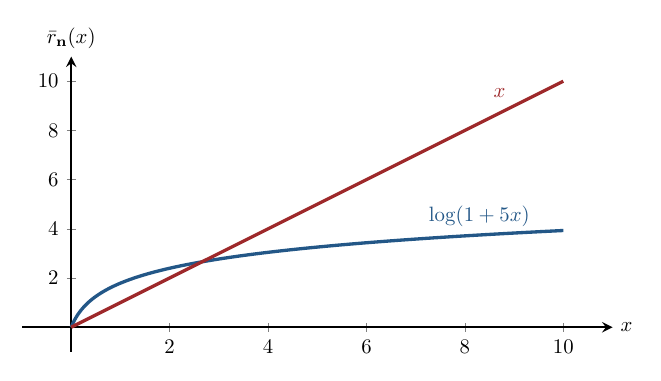}
\caption{Comparison between a convex regularizer (red) and a non-convex one (blue).
Both regularizers have a kink at $x=0$ that promotes sparsity. However, the convex regularizer
incurs a penalty that increases linearly with its argument, and thus creates a bias 
towards the origin. In contrast, the non-convex regularizer's penalty is close to constant
for large arguments, and thus causes a much smaller bias.}
\label{fig:regularization}
\vspace{-2mm}
\end{figure}

A remedy is to discard the debiasing stage, but this reintroduces the problem of bias that debiasing was supposed to solve. 
An alternative route for mitigating bias is to use non-convex regularization functions $r_\bmn$ that incur diminishing 
penalties on significant atoms while still promoting sparsity~\cite{zhang2010nearly}, see \fref{fig:regularization}.
Specifically, we want to use regularization functions
\begin{align}
    r_\bmn(x) &= r^{\tinysquare}_\bmn(x) + \bar{r}_\bmn(x) \\
    &= \gamma x^2 + \lambda_\bmn \log(1 + \eta x), 
\end{align}
where $r^{\tinysquare}_\bmn(x)=\gamma x^2$ is the quadratic part of $r_\bmn$ that prevents overfitting (see Footnote~\ref{footnote:tikhonov})
and $\bar{r}_\bmn(x)=\lambda_\bmn \log(1 + \eta x)$ is the non-quadratic part that promotes sparsity
(note that the weight~$\lambda_\bmn$ of the regularization is anisotropic, but its ``concaveness'' $\eta$ is isotropic).  
With this regularization, the pertinent optimization problem \eqref{eq:lr_optimization_problem} becomes
\begin{align}
    \{\hat\bma_\bmn\} \!=\! 
    \mathop{\textnormal{argmin}}_{\tilde{\bma}_{\bmn}:\,\bmn\in\setN}
    \big\|\bmy \!-\! \bsfS \tilde\bsfa \big\|^2_2  \!+\! \gamma \|\tilde\bsfa\|^2_2
    +\!\! \sum_{\bmn\in\setN}\! \lambda_\bmn \log(1 \!+\! \eta_\bmn \|\tilde{\bma}_\bmn\|_2). \label{eq:opt_nonconvex}
\end{align}

One thing to bear in mind, however, is that the proximal operator \eqref{eq:fbs_prox} becomes difficult to evaluate if
we try to reuse the same FBS optimization scheme as in \fref{sec:algorithm} while identifying $g$ from \eqref{eq:basic_fbs}
with the non-quadratic part of \eqref{eq:opt_nonconvex}.
In \fref{sec:algorithm}, the proximal operator amounted to simple group-wise soft-thresholding (see \eqref{eq:fbs_explicit_prox})
only because of the particular form of $g(\tilde\bsfa)=\lambda \sum_{\bmn\in\setN} \|\tilde{\bma}_\bmn\|$.
To preserve the algorithmic structure from \fref{sec:algorithm} and, in particular, the simplicity of the proximal step, 
we leverage the concept of multi-stage convex relaxation \cite{zhang2010analysis, candes2008enhancing}. 
In multi-stage convex relaxation, the idea is to (approximately) solve a non-convex optimization problem by solving a sequence of convex optimization problems. Specifically, to approximately solve~\eqref{eq:opt_nonconvex}, one iteratively solves
\begin{align}
    \{\hat\bma_\bmn^{(k)}\} \!=\! 
    \mathop{\textnormal{argmin}}_{\tilde{\bma}_{\bmn}:\,\bmn\in\setN}
    \big\|\bmy - \bsfS \tilde\bsfa \big\|^2_2  \!+ \gamma \|\tilde\bsfa\|^2_2
    +\!\! \sum_{\bmn\in\setN}\! \alpha_\bmn^{(k)} \|\tilde{\bma}_\bmn\|_2 \label{eq:iterative_nonconvex}
\end{align}
for $k=0,1,\dots,k_{\max}$, where $\alpha_\bmn^{(0)} \!\equiv\! 1$ for all $\bmn\!\in\!\setN$, and~where
\begin{align}
    \alpha_\bmn^{(k+1)} &= \frac{\partial}{\partial x} \bar{r}_\bmn(x)\Big|_{x=\|\hat\bma_\bmn^{(k)}\|_2} \label{eq:linearize}
    = \frac{\lambda_\bmn }{1+\eta_\bmn \,\big\|\hat\bma_\bmn^{(k)}\big\|_2}.
\end{align}
Note that the problem in \eqref{eq:iterative_nonconvex} retains the specific structure that leads to the soft-thresholding 
proximal operator \eqref{eq:fbs_explicit_prox}, except that the isotropic parameter $\lambda$ from \eqref{eq:fbs_explicit_prox} 
is now replaced with the anisotropic parameter $\alpha_\bmn^{(k)}$.
This multi-stage optimization strategy is motivated by the concept of majorization-minimization \cite{zhang2010analysis, candes2008enhancing}: 
The functions $\bar{r}_\bmn$ are concave and thus upper-bounded (or ``majorized'') by their tangents. 
One can therefore improve \eqref{eq:opt_nonconvex} from a current guess $\hat\bma_\bmn^{(k)}$ by minimizing a majorizing surrogate 
that linearizes the $\bar{r}_\bmn$ around $\hat\bma_\bmn^{(k)}$. 
This is exactly what \eqref{eq:iterative_nonconvex} does. 

We introduce an additional change with respect to this multi-stage convex relaxation scheme: According to theory, 
one would nest outer iterations over the different stages $k$ of \eqref{eq:iterative_nonconvex}, 
with inner iterations (over $t$ as in \fref{sec:algorithm}) that actually solve each of these stages. 
However, since these stages are only surrogate problems, it seems wasteful to solve them 
to completion. Instead, we therefore make only a single descent step per stage and re-majorize immediately. 

Ultimately, we end up with the procedure outlined in \fref{alg:dd_rkhs}, where all trainable parameters
are highlighted in red. 
Note that we make all parameters iteration dependent, meaning that, e.g., the regularization weights
$\lambda_\bmn^{(t)}$ or the concavity $\eta^{(t)}$ can change between iterations.
This means that there is no single, stable optimization problem being solved (or approximately solved)
by \fref{alg:dd_rkhs}. Instead, every algorithm iteration should be interpreted as a descent step 
on a dynamically evolving optimization problem.
The computational complexity of \fref{alg:dd_rkhs} is $O(t_{\max}R^3 |\setN| \log|\setN|)$.
The routine for learning the trainable parameters is described in \fref{sec:chest_baselines} below.

\begin{algorithm}[tp]
  \caption{Data-Driven RKHS Channel Estimation}
  \label{alg:dd_rkhs}
  \begin{algorithmic}[1]
    \Function{DD-RKHS-CHEST}{$\bmy$}
    \State $\tilde\bsfa^{(0)}=\herm\bsfS\bmy$
    \For{$t=0,1,\dots,t_{\max}-1$}
        \State \!\!$\breve\bsfa^{(t)} = \tilde\bsfa^{(t)} - \textcolor{myred}{\tau^{(t)}} \big(  \herm\bsfS(\bsfS\tilde\bsfa^{(t)} - \bmy) 
        + 2\textcolor{myred}{\gamma^{(t)}}\, \tilde\bsfa^{(t)} \big)$
        \State \!\!$\alpha_\bmn^{(t)} = \textcolor{myred}{\lambda^{(t)}_\bmn}/(1+\textcolor{myred}{\eta^{(t)}} \,\big\|\breve\bma_\bmn^{(k)}\big\|),\, \bmn\!\in\!\setN$
        \hfill \textcolor{myaltgray}{//\,re-majorization}
        \State \!\!$\tilde\bma_\bmn^{(t+1)} \!\!=\! (\breve\bma_\bmn^{(t)}\!\!/\|\breve\bma_\bmn^{(t)}\|_2) 
        \max\!\big\{\!\|\breve\bma_\bmn^{(t)}\|_2 - \textcolor{myred}{\tau^{(t)}} \alpha_\bmn^{(t)}\!\!, 0 \big\}, \bmn\!\in\!\setN$
    \EndFor
    \State \Return $\bsfS(\setF)\,\tilde\bsfa^{(t_{\max})}$
    \vspace{1mm}
    \EndFunction    
  \end{algorithmic}
\end{algorithm}

\section{Evaluation}
\label{sec:experiments}

\subsection{Simulation Settings}
\label{sec:settings}

\begin{figure}[tp]
\centering
\vspace{-2mm}
\includegraphics[width=0.95\columnwidth]{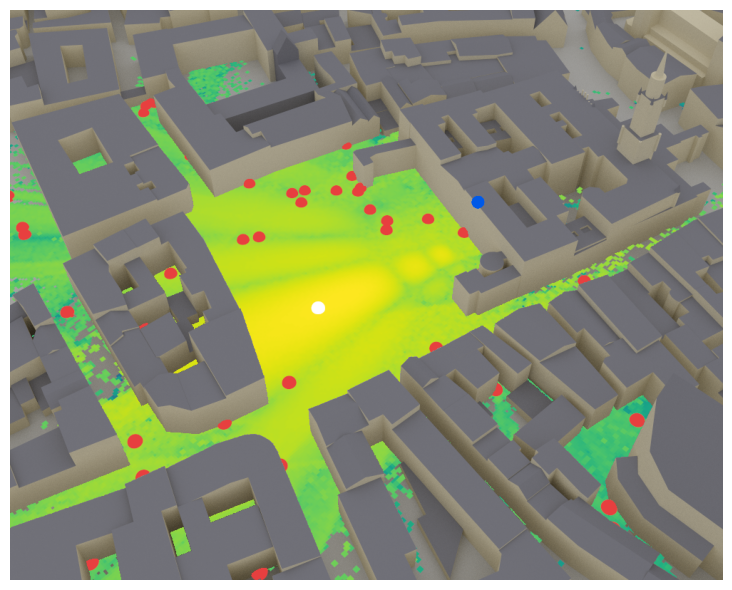}
\vspace{-2mm}
\caption{Visualization of our channel simulation environment in Sionna RT. 
The BS is placed at the blue dot, with its array facing
the white dot. The red dots correspond to randomly sampled user locations.}
\label{fig:sim_env}
\vspace{-2mm}
\end{figure}

We use channels generated with Sionna RT \cite{sionna, aoudia2025sionna} for our simulations. 
We consider a massive MIMO system with a multi-antenna BS overseeing the Marienhof in Munich, see \fref{fig:sim_env}: The blue dot indicates the BS location, the white dot indicates the point towards
which the BS is oriented, and the red dots indicate an exemplary set of randomly sampled user locations. 
The BS has 32 vertically polarized antennas with Sionna RT's \texttt{tr38901} antenna pattern 
\cite{aoudia2025sionna}. These antennas are arranged in a two-dimensional array with four rows 
and eight columns. The separation between adjacent rows corresponds to a full wavelength, and the 
separation between adjacent columns corresponds to half a wavelength. 
The UEs have a single, vertically polarized antenna with isotropic radiation. 
The UE locations are randomly sampled from a horizontal plane 1.5m above ground, subject to the following 
constraints: the path gain is at least -140dB and at most -80dB, and the distance to the BS is at most 250m. 
The channels are ray-traced with refraction and diffuse reflection enabled, where the scattering coefficient
of each material in the scene is set to $0.15$, with a directive scattering pattern with directivity
$\alpha_R=10$ \cite{degli2007measurement}.

We consider an OFDM system with a bandwidth of 100\,MHz centered around a carrier frequency 3.5\,GHz. 
The subcarrier spacing is 30\,kHz and the total number of subcarriers is $\Nsc=3276$. Pilots are transmitted
on every fourth subcarrier, so that there are $\Npil=819$ pilot-carrying subcarriers. 
The UEs transmit with a maximum power of 23\,dBm, and the BS has a noise power of -89\,dBm.
At maximum transmit power, these parameters would entail an SNR that ranges
from -28\,dB (at a channel gain of -140) to 32\,dB. However, we assume that the UEs use a power 
control strategy that caps the receive power at the BS at -68dBm, which limits the SNR at the BS to 21\,dB. 
Since we are using Monte Carlo simulations (see below), we bin the channels into channel gain bins of 3\,dB width
and we round all channel gains within a bin to the bin's center (otherwise, every channel would have
a distinct SNR, which would make it impossible to statistically average the performance over different channels for a given SNR). 
We assume that the BS uses automatic gain control (AGC) at the input such that the channel gain \emph{after}
AGC is equal to one.
\vspace{-1mm}

\subsection{Channel Estimators and Baselines}
\label{sec:chest_baselines}

We study the performance of our proposed channel estimators with the following settings:

\vspace{1mm}
\subsubsection*{RKHS Channel Estimator (RKHS)}
We consider the RKHS channel estimator from \fref{alg:rkhs} with Nyquist sampling, 
$t_{\max}=30$ iterations with stepsize $\tau^{(t)}\equiv 1.0$ for sparse optimization, 
$t'_{\max}=30$ iterations with stepsize $\tau^{(t)}\equiv 1.0$ for debiasing, 
and isotropic regularization constant $\lambda=500\sqrt{\No}$. 
The dimension of the low-rank approximation is $R=3$.
Variations of these parameters are studied in \fref{app:parameter_variations}.

\vspace{1mm}
\subsubsection*{Data-Driven RKHS Channel Estimator (DD-RKHS)}
For the data-driven RKHS channel estimator from \fref{alg:rkhs}, we use $t_{\max}=5$ algorithm iterations.
To learn the trainable parameters of \fref{alg:dd_rkhs}, we create a training set of channels
from the same environment as will be used in the evaluation, see \fref{sec:settings}.\footnote{The training 
set consists of channels that are distinct from those in the test set.}
We generate a training set of approximately 12'000 channels, ensuring that the entire range of channel gains is adequately represented. 
We learn different parameter sets for different channel gains. 
For training, we use the log-NMSE as loss function, and we use Adam \cite{kingma2014adam}
with AMSGrad \cite{reddi2019convergence}, a batch size of 40, and a learning rate of $0.01$.

Note that, since DD-RKHS's computational complexity per iteration is comparable to RKHS's computational 
complexity per iteration (both for sparse optimization and debiasing), the DD-RKHS channel estimator has about 
twelve times lower computational complexity than the RKHS channel estimator. 

\vspace{1mm}
We compare our estimators with the following baselines: 

\vspace{1mm}
\subsubsection*{LS Channel Estimator (LS)}
This channel estimator performs least squares estimation of the pilot-carrying subcarriers
and subsequent nearest-neighbor interpolation in between subcarriers. 

\vspace{1mm}
\subsubsection*{Empirical LMMSE Channel Estimator (Emp.\ LMMSE)}
This channel estimator performs least squares estimation of the pilot-carrying subcarriers. 
Using Sionna's \texttt{LMMSEInterpolator}, it then interpolates in between subcarriers with LMMSE 
interpolation (which also smoothes the estimates of the pilot-carrying subcarriers) and subsequently 
performs spatial smoothing with LMMSE filtering. 
The required covariance matrices are estimated as the empirical averages of the
spectral and spatial correlation matrices of 5000 randomly sampled channels from the same environment 
as the evaluation channels. 

\vspace{1mm}
\subsubsection*{Genie LMMSE Channel Estimator (Genie LMMSE)}
This channel estimator is provided (by a genie) with the ground-truth $\Nsc\Nrow\Ncol\times\Nsc\Nrow\Ncol$ covariance matrix 
of a given vectorized channel, which it uses for LMMSE interpolation and smoothing of the least-squares initial channel estimate.
We consider this estimator as a practical upper bound on the attainable performance.\footnote{Strictly speaking, it is not
a bound on mean-squared error performance: since there is only a finite number of paths in each channel, 
the channels are only approximately Gaussian, and the LMMSE estimator is only an approximation to the minimum 
mean squared error (MMSE) estimator.}

\subsection{Normalized Mean Squared Error (NMSE) Performance}
%
\begin{figure}[tp]
\centering
\vspace{-0.1mm}
\includegraphics[scale=0.8]{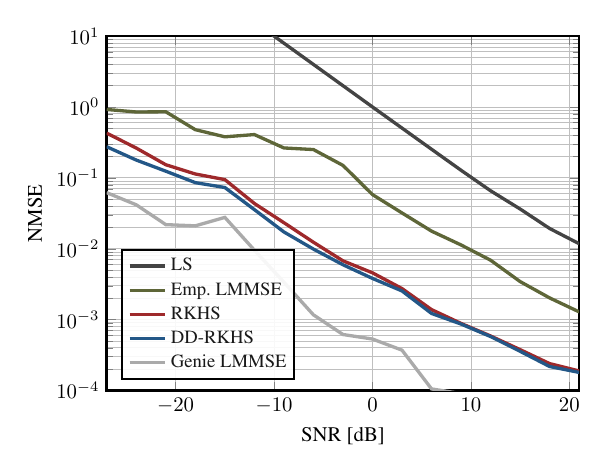}
\caption{Accuracy of different channel estimators, measured in normalized mean squared error (NMSE) as a function of signal-to-noise ratio (SNR).}
\label{fig:sim_comparison}
\end{figure}

We start by evaluating the channel estimation performance using the normalized mean squared error (NMSE), 
which is computed as a function of the SNR as 
\begin{align}
    \textit{NMSE}(\textit{SNR}) = \Ex{}{\frac{\|\bmh-\hat\bmh\|_2^2}{\|\bmh\|_2^2}}(\textit{SNR}),
\end{align}
where $\bmh, \hat\bmh\in\opC^{\Nsc\Nrow\Ncol}$ are the vectorized ground truth channel and channel estimate
over all subcarriers and all antennas, and where the expectation is over channel and noise realizations. 
We approximate the NMSE as a function of the SNR with Monte Carlo simulations. 

\fref{fig:sim_comparison} shows the NMSE as a function of the SNR for the different channel estimators 
from \fref{sec:chest_baselines}. 
(The ruggedness of the Emp. LMMSE and Genie LMMSE curves is due to their high computational complexity, which only allowed
for a small number of Monte Carlo trials.)
RKHS and DD-RKHS outperform LS by 26\,dB and 27\,dB, respectively, and LMMSE by 15\,dB and 16\,dB, respectively, 
in the SNR required to achieve an NMSE of 1\%. Compared to the Genie LMMSE bound, they only lose 6\,dB and 7\,dB in SNR, 
respectively. 
Even though DD-RKHS has an order of magnitude lower computational complexity than RKHS, it outperforms vanilla RKHS
by between 1\,dB and 3\,dB in SNR when the SNR is below 0\,dB, and between 0.1\,dB and 1\,dB when the SNR is above 0\,dB. 
These results demonstrate the efficacy of our channel estimator based on RKHS regularization 
as well as of our data-driven extensions.

\subsection{Achievable Rate Performance}
\label{sec:achievable_rate}

Since the NMSE evaluates the accuracy of channel estimation but does not tell how gains (or losses) in accuracy 
bear on the communication performance, we now consider a second, achievable-rate based figure of merit.
We assume a single-user downlink where the BS uses the obtained channel estimate for downlink beamforming.
The following downlink-rate is achievable~\cite{shannon48a} with maximum ratio combining:
\begin{align}
    \mathsf{R} = \frac{1}{\mathsf{T}_s} \sum_{f\in\setF}\log_2\bigg(1+\frac{\herm{\hat\bmh_f}\bmh_f}{\|\hat\bmh_f\|_2}\frac{P_{\text{BS}}}{N_{0,\text{UE}}}\bigg), \label{eq:achievable_rate}
\end{align}
where $\mathsf{T}_s$ is the OFDM symbol duration (which we take to be the inverse of the subcarrier spacing, 
i.e., we neglect impact of the cyclic prefix on symbol duration),
$P_{\text{BS}}$ is the transmit power at the BS, $N_{0,\text{UE}}$ is the noise power at the UE, 
and $\bmh_f, \hat\bmh_f\in\opC^{\Nrow\Ncol}$ are the vectorized channel and channel estimate (over all antennas) 
on the subcarrier with frequency~$f$. Note that we assume in \eqref{eq:achievable_rate} that the BS transmits
with identical power on all subcarriers (rather than using water-filling \cite{gallager1968information}) 
because of potential regulatory limits on power spectral density. 
We assume a BS transmit energy of 49\,dBm and a UE noise power of -85\,dBm. 
Note that, by channel reciprocity, the downlink SNR is therefore equal to the uplink SNR plus 22\,dB 
(26\,dB higher transmit power compared to 4\,dB higher noise power) plus the beamforming gain:
\begin{align*}
    [\text{DL SNR}]_{\text{dB}} = [\text{UL SNR}]_{\text{dB}} + 22\,\text{dB} + [\text{beamforming gain}]_{\text{dB}}
\end{align*}

\begin{figure}[tp]
\centering
\includegraphics[scale=0.8]{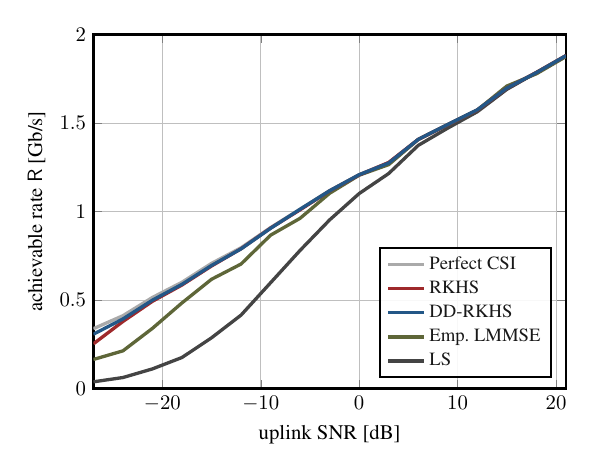}
\caption{Single-user downlink achievable rates when channel estimates of different estimators are used for 
maximum ratio combining (MRC) precoding.}
\label{fig:sim_achievable_rates}
\end{figure}

\fref{fig:sim_achievable_rates} shows the average (over channel and noise realizations) 
achievable rate as a function of the uplink SNR for 
the different channel estimators from \fref{sec:chest_baselines}.\footnote{Note that we have replaced the Genie LMMSE
channel estimation bound with a bound that uses perfect channel state information 
(Perfect CSI).}
The results show that all channel estimators achieve the performance of the upper bound at high SNR, 
but that there are significant differences at low SNR: At -21dB uplink SNR, 
LS channel estimation attains 22\% of the rate achievable with perfect channel state information (CSI), 
empirical LMMSE channel estimation attains 66\%, and the RKHS and DD-RKHS channel estimators
attain 96\% and 97\%, respectively. 
At -27dB uplink SNR, these channel estimators attain 12\%, 49\%, 75\%, and 91\%, respectively, 
of the rate achievable with perfect CSI.

\section{Conclusions}/
\label{sec:conclusions}
We have presented a new model-based channel estimator that leverages the band-sparsity of wireless channels
in delay-beamspace using regularization in an RKHS. 
To solve the resulting optimization problem, we have derived an efficient iterative algorithm with the help of 
a novel representer theorem, dimensionality-reduction techniques, and by exploiting the RKHS's modulation structure.
We have also proposed a data-driven extension with additional hyperparameters in the form of anisotropic
non-convex regularization. These hyperparameters are learned in supervised manner. 
Simulations show that both estimators significantly outperform linear estimators, and that the data-driven 
version slightly outperforms its classical counterpart at significantly lower computational complexity.


\appendices

\makeatletter
\renewcommand{\thesubsection}{\Alph{section}-\Roman{subsection}}    
\renewcommand{\thesubsectiondis}{\Roman{subsection}.} 
\makeatother

\section{Variation of RKHS Parameters}
\label{app:parameter_variations}

In this appendix, we investigate individual variations in the settings of the RKHS channel estimator 
compared to the ones described in \fref{sec:experiments}. We only show the effect on NMSE performance. 
In each experiment, the remaining parameters remain as described in \fref{sec:chest_baselines}.

\begin{figure}[tp]
\centering
\includegraphics[width=0.9\columnwidth]{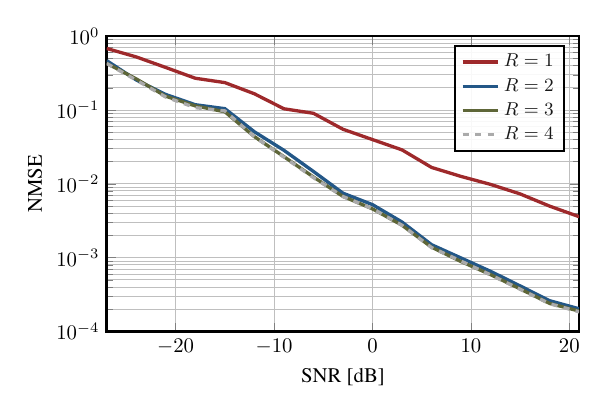}
\vspace{-5mm}
\caption{NMSE performance for different low-rank dimensions $R$.}
\label{fig:sim_low_rank}
\end{figure}

In \fref{fig:sim_low_rank}, the dimension $R$ of the low-rank approximation is varied.
The results show that $R=1$ results in significant performance degradation, $R=2$ deteriorates only marginally, 
and there is no difference at all between $R=3$ and $R=4$. This suggests that ideal performance is obtained
already for $R=3$.

\begin{figure}[tp]
\centering
\vspace{-5.5mm}
\includegraphics[width=0.9\columnwidth]{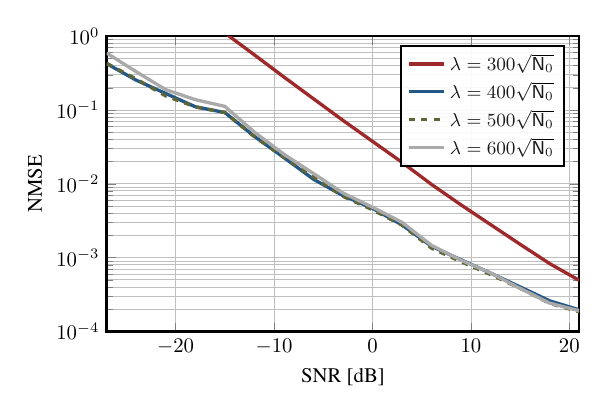}
\vspace{-5mm}
\caption{NMSE performance for different regularization constants $\lambda$.}
\label{fig:sim_reg}
\end{figure}

In \fref{fig:sim_reg}, the proportionality factor between the regularization constant $\lambda$ and the noise
intensity $\sqrt{\No}$ is varied. The results show that a proportionality factor between $400$ and $500$ appears 
to be optimal for all SNRs.

\begin{figure}[tp]
\centering
\includegraphics[width=0.9\columnwidth]{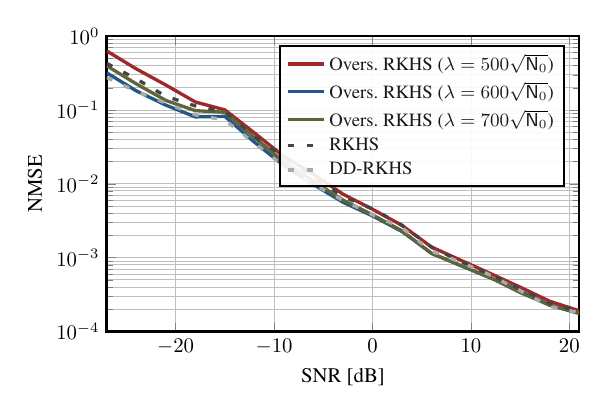}
\vspace{-5mm}
\caption{NMSE performance with oversampling by a factor of $2$.}
\label{fig:sim_oversampling}
\end{figure}

In \fref{fig:sim_oversampling}, we study the performance when using oversampling with an 
oversampling factor of $2$ (note that this increases the computational complexity by a factor of $2^3=8$).
Note that oversampling changes the sparsity pattern and therefore affects the optimal value of $\lambda$. 
For optimally tuned $\lambda$, oversampling increases the performance of RKHS to roughly match the performance 
of Nyquist-sampled DD-RKHS (but it has \emph{much} higher computational complexity).

\begin{figure}[tp]
\centering
\vspace{-5mm}
\includegraphics[width=0.9\columnwidth]{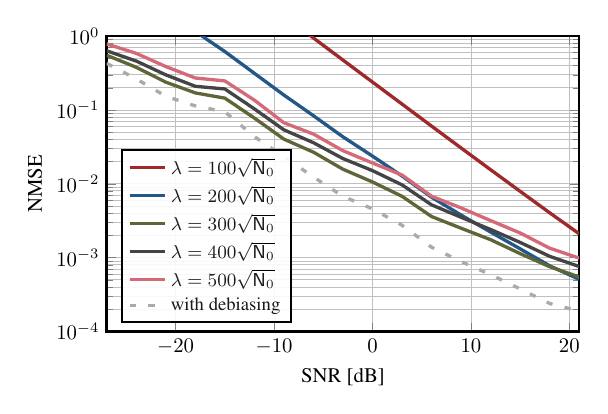}
\vspace{-5mm}
\caption{NMSE performance without debiasing.}
\label{fig:sim_no_debiasing}
\end{figure}

In \fref{fig:sim_no_debiasing}, we study the performance without debiasing, for different values of $\lambda$. 
The optimal $\lambda$ with debiasing ($\lambda=400\sqrt{\No}$) now performs worse
than $\lambda=300\sqrt{\No}$ because it introduces a stronger bias. 
Lowering $\lambda$ to $\lambda=200\sqrt{\No}$ also decreases performance
because sparsity is no longer properly exploited. Even at the optimal tradeoff ($\lambda=300\sqrt{\No}$),
the absence of debiasing decreases performance significantly. 

\begin{figure*}[tp]
\centering
\subfigure[varying stepsizes/number of iterations for sparse optimization~(Alg.\ref{alg:rkhs})]{
    \includegraphics[width=0.9\columnwidth]{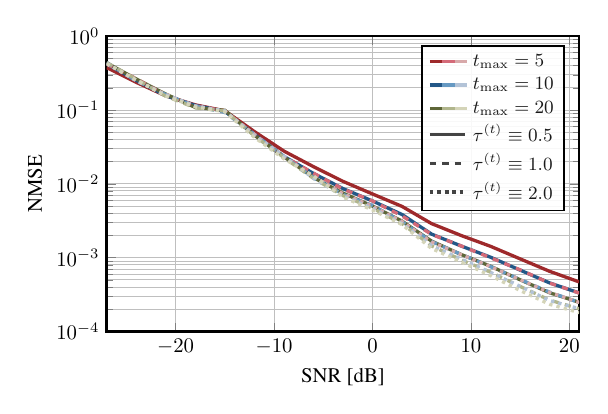}\label{fig:sparse_it}
    \hspace{6.5mm}
    
}
\subfigure[varying stepsizes/number of iterations for debiasing (Alg.\,\ref{alg:rkhs})]{
    \hspace{2mm}
    \includegraphics[width=0.9\columnwidth]{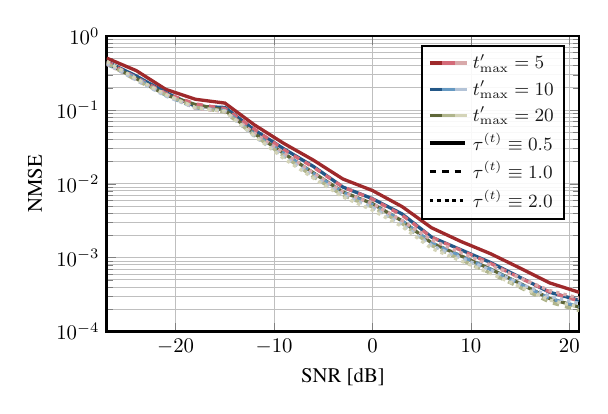}\label{fig:debiasing_it}
}
\caption{NMSE performance with different stepsizes/number of iterations in either of the algorithm stages
(the other stage uses $30$ iterations with $\tau^{(t)}=1.0$).}
\label{fig:sim_num_it}
\end{figure*}

In \fref{fig:sim_num_it}, we consider the performance with fewer iterations 
(either in the sparse optimization stage, \fref{fig:sparse_it}, or in the debiasing stage, \fref{fig:debiasing_it})
and with different stepsizes.
The results show that, regardless of the stepsize (stepsizes larger than $\tau^{(t)}=2$ tend to diverge), 
simply reducing the number of iterations decreases performance.\footnote{Note that RKHS 
uses the same stepsize in all iterations, while DD-RKHS can use different stepsizes in different iterations.}

\begin{figure}[tp]
\centering
\vspace{-5mm}
\includegraphics[width=0.9\columnwidth]{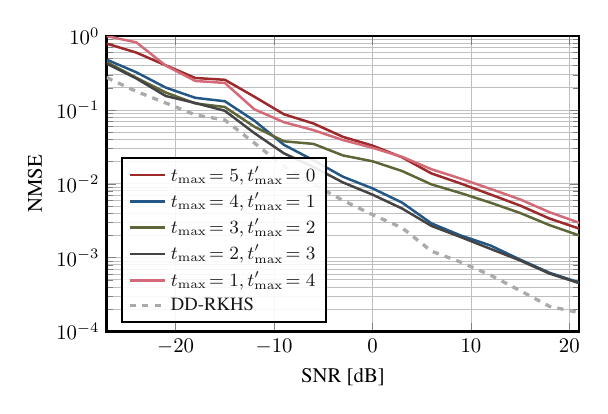}
\vspace{-5mm}
\caption{NMSE performance when RKHS is constrained to the same computational complexity as DD-RKHS.}
\vspace{-1mm}
\label{fig:sim_efficient}
\end{figure}

In \fref{fig:sim_efficient}, we study the performance of RKHS when constrained to the same computational 
complexity as DD-RKHS (i.e., to the same number of total iterations $t_{\max}+t'_{\max}=5$).
The results show that, regardless of how the iterations are distributed to the stages, RKHS performs significantly 
worse than DD-RKHS at identical computational complexity.

\section{Proofs}
\subsection{Proof of \fref{thm:representer}}
\label{app:thm_proof}

Let $\tilde{h}=\sum_{\bmn\in\setN}\tilde{h}_\bmn$ with $\tilde{h}_\bmn\in\setR_\bmn(\setZ)$. 
Moreover, let $V_\bmn = \text{span}\big(\{K_\bmn(\bmz_\bmm,\cdot)\}_{\bmm\in\setM}\big)\subset\setR_\bmn(\setZ)$, 
and let $P_{V_\bmn}$ be the orthogonal projection onto $V_\bmn$.
At the points $\bmz_\bmm$, we have
\begin{align}
    \tilde{h}_\bmn(\bmz_\bmm) &= \big\langle \tilde{h}_\bmn, K_n(\bmz_\bmm,\cdot) \big\rangle \label{eq:use_reproducing_property} \\
    &= \big\langle \tilde{h}_\bmn, P_{V_\bmn}K_\bmn(\bmz_\bmm,\cdot) \big\rangle \label{eq:proj_kernel} \\
    &= \big\langle P_{V_\bmn}\tilde{h}_\bmn, K_\bmn(\bmz_\bmm,\cdot) \big\rangle \label{eq:adjoint_proj} \\
    &= \big(P_{V_\bmn}\tilde{h}_\bmn\big)(\bmz_\bmm), \label{eq:use_reproducing_property2}
\end{align}
where \eqref{eq:use_reproducing_property} follows from the reproducing property of RKHS, 
\eqref{eq:proj_kernel} follows since $K_\bmn(\bmz_\bmm,\cdot)\in V_\bmn$, 
\eqref{eq:adjoint_proj} follows since $P_{V_\bmn}$ is an orthogonal projection, 
and \eqref{eq:use_reproducing_property2} follows again from the reproducing property, 
since $P_{V_\bmn}\tilde{h}_\bmn \in \setR_\bmn(\setZ)$.

For the first term of the objective in \eqref{eq:opt_prob}, we therefore obtain
\begin{align}
\sum_{\bmm\in\setM}\Big|y_\bmm - \!\sum_{\bmn\in\setN}\tilde{h}_\bmn(\bmz_\bmm)\Big|^2  
&\!=\!\sum_{\bmm\in\setM}\Big|y_\bmm - \!\sum_{\bmn\in\setN}P_{V_\bmn} \tilde{h}_\bmn(\bmz_\bmm)\Big|^2. \label{eq:term1_ineq}
\end{align}
Moreover, since projections are contractive and the $r_n$ are increasing, it follows with regards to the 
second term of the objective in \eqref{eq:opt_prob} that
\begin{align}
    \sum_{\bmn\in\setN} r_\bmn\big(\|\tilde{h}_\bmn\|\big)
    &\geq \sum_{\bmn\in\setN} r_\bmn\big(\|P_{V_\bmn} \tilde{h}_\bmn\|\big) \label{eq:term2_ineq}
\end{align}
where equality holds if $\tilde{h}_\bmn\in V_\bmn$ for all  $\bmn\in\setN$.
By combining \eqref{eq:term1_ineq} and \eqref{eq:term2_ineq}, we get
\begin{align}
    &\sum_{\bmm\in\setM}\Big|y_\bmm - \sum_{\bmn\in\setN}P_{V_\bmn} \tilde{h}_\bmn(\bmz_\bmm)\Big|^2 
    + \sum_{\bmn\in\setN} r_\bmn\big(\|P_{V_\bmn} \tilde{h}_\bmn\|\big) \nonumber\\
    &\leq \sum_{\bmm\in\setM}\Big|y_\bmm - \sum_{\bmn\in\setN}\tilde{h}_\bmn(\bmz_\bmm)\Big|^2 + \sum_{\bmn\in\setN} r_\bmn\big(\|\tilde{h}_\bmn\|\big)
\end{align}
with equality if $\tilde{h}_\bmn\!\in\! V_\bmn$ for all  $\bmn\!\in\!\setN$.
It follows that~there~exists a minimizing $\tilde{h}=\sum_{\bmn\in\setN}\tilde{h}_\bmn$ of \eqref{eq:opt_prob}
that satisfies \mbox{$\tilde{h}_\bmn\!\in\! V_\bmn$} for all  $\bmn\in\setN$.
These $\tilde{h}_\bmn$ admit a representation of the form 
\begin{align}
    \tilde{h}_\bmn(\cdot) = \sum_{\bmm\in\setM} \tilde{c}_{\bmn,\bmm}\, K_\bmn(\bmz_\bmm,\cdot). 
\end{align}
To find the minimizing coefficients $\tilde{c}_{\bmn,\bmm}$, we plug this representation into the objective of \eqref{eq:opt_prob}. 
For the first term, we obtain
\begin{align}
    \sum_{\bmm\in\setM}\Big|y_\bmm - \sum_{\bmn\in\setN}\sum_{\bmm'\in\setM} \tilde{c}_{\bmn,\bmm'} K_\bmn(\bmz_{\bmm'},\bmz_\bmm)\Big|^2, 
\end{align}
and for the second term, we obtain 
\begin{align}
    &\!\!\!\!\sum_{\bmn\in\setN} r_\bmn\Big(\Big\| \sum_{\bmm\in\setM} \tilde{c}_{\bmn,\bmm} K_\bmn(\bmz_\bmm,\cdot) \Big\|_2\Big) \nonumber\\
    &\!\!\!\!= \sum_{\bmn\in\setN} r_\bmn\Big(\! \Big\langle\! \sum_{\bmm\in\setM} \tilde{c}_{\bmn,\bmm} K_\bmn(\bmz_\bmm,\cdot),\!\! \sum_{\bmm'\in\setM} \tilde{c}_{\bmn,\bmm'} K_\bmn(\bmz_{\bmm'},\cdot) \Big\rangle^{\!\frac12} \Big) \\
    &\!\!\!\!= \!\sum_{\bmn\in\setN}\! r_\bmn\Big(\!(\!\sum_{\bmm\in\setM} \sum_{\bmm'\in\setM}\! \tilde{c}_{\bmn,\bmm} \tilde{c}_{\bmn,\bmm'}^\ast 
    \big\langle K_\bmn(\bmz_\bmm,\cdot), K_\bmn(\bmz_{\bmm'},\cdot) \big\rangle)^{\frac12}\Big) \\
    &\!\!\!\!= \sum_{\bmn\in\setN} r_\bmn\Big((\sum_{\bmm\in\setM} \sum_{\bmm'\in\setM} \tilde{c}_{\bmn,\bmm} \tilde{c}_{\bmn,\bmm'}^\ast K_\bmn(\bmz_\bmm, \bmz_{\bmm'}))^{\frac12}\Big). 
\end{align}
It follows that the coefficients $\tilde{c}_{\bmn,\bmm}$ that minimize (via \eqref{eq:rep_sol}) the objective in \eqref{eq:opt_prob}
are found by minimizing \eqref{eq:rep_coeff}.
\hfill$\blacksquare$

\subsection{Proof of \fref{lem:phase}}
\label{app:lem_proof}

Let $(\bK_{n_2}^{\setH}\bmv_{n_2})[m]$ and $(\bK^{\setH}\bmv)[m]$ be the $m$th entries of $\bK_{n_2}^{\setH}\bmv_{n_2}$ 
and $\bK^{\setH}\bmv$, respectively, and let $\tp{\bmk_{n_2}}[m]$ and $\tp{\bmk}[m]$ be the $m$th rows of $\bK_{n_2}^{\setH}$ 
and $\bK^{\setH}$, respectively. We have
\begin{align}
    \tp{\bmk_{n_2}}[m] = \exp\!\bigg(\!2\pi i x^{(\text{r})}_m \frac{(n_2-\frac{N_2+1}{2})\Phi^{(\text{h})}}{N_2} \bigg) (\tp{\bmk}[m] \odot \herm{(\bmf_{n_2}^{\setH})})
\end{align}
and therefore
\begin{align}
&(\bK_{n_2}^{\setH}\bmv_{n_2})[m] \nonumber \\
&= \tp{\bmk_{n_2}}[m] \bmv_{n_2} \\
&= \exp\!\bigg(\!2\pi i x^{(\text{r})}_m \frac{(n_2-\frac{N_2+1}{2})\Phi^{(\text{h})}}{N_2} \bigg) (\tp{\bmk}[m] \odot \herm{(\bmf_{n_2}^{\setH})}) \bmv_{n_2} \\
&= \exp\!\bigg(\!2\pi i x^{(\text{r})}_m \frac{(n_2-\frac{N_2+1}{2})\Phi^{(\text{h})}}{N_2} \bigg) \tp{\bmk}[m] ((\bmf_{n_2}^{\setH})^\ast \odot \bmv_{n_2}) \\
&= \exp\!\bigg(\!2\pi i x^{(\text{r})}_m \frac{(n_2-\frac{N_2+1}{2})\Phi^{(\text{h})}}{N_2} \bigg) \tp{\bmk}[m] \bmv \\
&= \exp\!\bigg(\!2\pi i x^{(\text{r})}_m \frac{(n_2-\frac{N_2+1}{2})\Phi^{(\text{h})}}{N_2} \bigg) (\bK^{\setH}\bmv)[m].
\end{align}
It follows that 
\begin{align}
    \bK_{n_2}^{\setH}\bmv_{n_2} &= \bmf_{n_2}^{\setH} \odot (\bK^{\setH}\bmv) \\
    &= \bmf_{n_2} \odot (\lambda \bmv) \\
    &= \lambda \bmv_{n_2}, 
\end{align}
which proves that $\bmv_{n_2}$ is an eigenvector of $\bK_{n_2}^{\setH}$ with corresponding eigenvalue $\lambda$.
\hfill$\blacksquare$

\balance


\end{document}